\documentclass[aps,prl,twocolumn,amsmath,amssymb,superscriptaddress,longbibliography]{revtex4-2}

\usepackage{graphicx}
\usepackage{color}
\usepackage[hidelinks]{hyperref}
\usepackage{orcidlink}
\usepackage{amsmath}
\usepackage{amsthm}
\usepackage{physics}
\usepackage{amsfonts}
\usepackage{times,txfonts}
\usepackage[normalem]{ulem}

\hypersetup{colorlinks=true, linkcolor=black, citecolor=black, urlcolor=blue}

\begin{document}

\author{Christoph Hotter\,\orcidlink{0009-0003-3854-0264}}
\affiliation{Niels Bohr Institute, University of Copenhagen, Blegdamsvej 17, Copenhagen DK-2100, Denmark} 

\author{Adam Miranowicz\,\orcidlink{0000-0002-8222-9268}}
\affiliation{Institute of Spintronics and Quantum Information, Faculty of Physics and Astronomy, Adam Mickiewicz University, PL-61-614 Pozna\'n, Poland}

\author{Karol Gietka\,\orcidlink{0000-0001-7700-3208}}
\email[]{karol.gietka@uibk.ac.at}
\affiliation{Institut f\"ur Theoretische Physik, Universit\"at Innsbruck, Technikerstra{\ss}e\,21a, A-6020 Innsbruck, Austria} 

\title{Quantum Metrology in the Ultrastrong Coupling Regime of
Light-Matter Interactions: \\ Leveraging Virtual Excitations without
Extracting Them}


\begin{abstract}
Virtual excitations, inherent to ultrastrongly coupled light-matter systems, induce measurable modifications in system properties, offering a novel resource for quantum technologies. In this work, we demonstrate how these virtual excitations and their correlations can be harnessed to enhance precision measurements, without the need to extract them. Building on the paradigmatic Dicke model, which describes the interaction between an ensemble of two-level atoms and a single radiation mode, we propose a method to harness hybridized light-matter modes whose renormalized frequencies encode the effects of virtual excitations for quantum metrology. Remarkably, we find that for a fixed squeezing parameter $\xi$, exploiting virtual squeezing through oscillator frequency shifts yields a quadratic enhancement in estimation precision—scaling as $\exp(4\xi)$—compared to the conventional $\exp(2\xi)$ scaling of real squeezed states. These results show that virtual excitations, though unobservable, can drive metrological performance beyond the standard quantum limit. Our approach establishes a broadly applicable framework for high-precision measurements across a wide class of ultrastrongly coupled quantum systems.

\end{abstract}
\date{\today}
\maketitle


\emph{Introduction.}---
Recent theoretical and experimental advances in understanding and controlling light-matter interactions have brought us closer to exploring the physics of ultrastrong coupling, a regime in which the light-matter coupling strength becomes comparable to the system's transition frequencies~\cite{Nori2019USCreview,solano2019rmp,QIN20241}. While the weak and strong coupling regimes have already driven significant progress in quantum technologies~\cite{quantumtechnologies2003Milburn,QT2015Schmiedmayer}, the ultrastrong coupling regime remains largely unexplored. This regime gives rise to counterintuitive and intriguing phenomena~\cite{bastard2005vacuumproperties,Ciuti2006inputoutputUSC,Ciuti2009signaturesUSC,Todorov2010USCpolaritondots,Nori2010qubitoscillatorUSC,blais2011dissipationUSC,savasta2018dissipationUSC,Teufel2019Ulstrastrongmechanicalcavity,uros2023dissipativephasetrans,gietks2023USquezingQRM,stassi2023unvelingveiling,UrosDelic2024OPtomochenicalUSC,chen2024suppressedenergyrelaxationquantum} including highly entangled ground states. However, this entanglement is purely virtual, meaning that it does not correspond to directly measurable correlations between atoms and photons. Instead, it manifests indirectly through modifications to observable system properties, such as non-linear frequency shifts of normal modes, which depend on the system's parameters~\cite{Cao_2011Noriassymetry,gietka2024vacuumrabisplittingmanifestation}.  If such modifications to physical excitations can be precisely controlled, they could potentially enable novel applications in quantum technologies.

In this article, we propose a method to leverage virtual excitations in ultrastrongly coupled systems for precision measurements. We consider an ensemble of atoms ultrastrongly coupled to a single mode of radiation, modeled by the paradigmatic Dicke model. Our approach involves coupling the to-be-measured field with a control system to form hybridized modes and performing all measurements on these hybrid modes instead of extracting virtual excitations. The key aspect {is} 
the non-trivial dependence of the hybrid mode frequency on the system's parameters {as well as on virtual excitations}. This work introduces a novel direction in quantum metrology, demonstrating how unique properties of ultrastrong coupling can be harnessed for practical applications. 


\emph{Dicke Model in the ultrastrong coupling regime.}---The Dicke model describes an ensemble of two-level systems collectively coupled to a single mode of electromagnetic radiation~\cite{garraway2011dickerevisisted}. Typically, the coupling strength between light and matter is much smaller than the atomic transition frequency, allowing the use of the rotating wave approximation~\cite{helmut2013rmp}. This approximation eliminates counter-rotating terms, reducing the Dicke model to the Tavis-Cummings model. In the ultrastrong coupling regime, where the coupling strength is comparable to the characteristic frequencies of the system, the rotating wave approximation breaks down, and counter-rotating terms must be included~\cite{emarybrandes2003dickechaosphase}. This inclusion fundamentally alters the system's properties. To understand this more clearly, we now consider the Dicke model Hamiltonian ($\hbar = 1$)
\begin{align}
    \hat{H} = \omega \hat{a}^\dagger \hat{a} + \Omega \hat{S}_z + \frac{g}{\sqrt{N}} \left( \hat{a} + \hat{a}^\dagger \right) \hat{S}_x,
\end{align}
where $\hat{a}$ ($\hat{a}^\dagger$) is the annihilation (creation) operator for a photon with frequency $\omega$, $\hat{S}_i = \sum_{n=1}^N \hat{\sigma}_i^n / 2$---where $\hat\sigma_i^n$ is the $i$th ($i=x,y,z$) Pauli matrix describing the $n$th atom---are collective spin operators representing $N$ two-level systems with transition frequency $\Omega$, and $g$ is the vacuum Rabi coupling strength. Close to the resonance, $\omega \approx \Omega$, the ultrastrong coupling regime is characterized by $g \lesssim \omega \approx \Omega$, approaching the critical coupling strength $g_c = \sqrt{\omega \Omega}$ for the superradiant phase transition~\cite{Larson_2017}.

For large ensembles ($N \gg 1$), the collective spin operators can be approximated by harmonic oscillator operators, which becomes exact in the thermodynamic limit ($N \to \infty$). Under this approximation, the Dicke Hamiltonian takes the form 
\begin{align}\label{eq:HDickeapprox}
    \hat{H} = \omega \hat{a}^\dagger \hat{a} + \Omega \hat{b}^\dagger \hat{b} + \frac{g}{2} \left( \hat{a} + \hat{a}^\dagger \right) \left( \hat{b} + \hat{b}^\dagger \right),
\end{align}
where $\hat{b}$ and $\hat{b}^\dagger$ represent the collective atomic mode. The ground state of this Hamiltonian is a two-mode squeezed vacuum state, which under the resonance condition ($\omega = \Omega$), can be written as~\cite{zhou2HO2020}
\begin{align}
    |\xi_-, \xi_+\rangle = \exp\left[{\frac{\xi_-}{2} \left(\hat{c}^2 - \hat{c}^{\dagger 2}\right)}\right] \exp\left[{\frac{\xi_+}{2} \left(\hat{d}^2 - \hat{d}^{\dagger 2}\right)}\right] |0_a, 0_b\rangle,
\end{align}
where $\hat c = (\hat a - \hat b)/\sqrt{2}$ and $\hat d = (\hat a + \hat b)/\sqrt{2}$ are the light-matter modes that are being squeezed, $\xi_\pm =\frac{1}{4}\log(1 \pm g/g_c)$ 
are the corresponding squeezing parameters (see Ref.~\cite{emarybrandes2003dickechaosphase} for a general {case}
), $|0_{a}\rangle$ and $|0_{b}\rangle$ {represent the vacuum states} 
of mode $\hat a$ and $\hat b$, respectively. 

From the viewpoint of the noninteracting modes $\hat a$ and $\hat b$, such a ground state contains excitations
\begin{align}
    \langle \hat a^\dagger\hat a \rangle = \langle \hat b^\dagger\hat b \rangle = \frac{1}{2}\left(\sinh^2 \xi_- + \sinh^2\xi_+ \right),
\end{align}
in a correlated form of two-mode squeezing between light ($\hat a$) and matter ($\hat b$). However, these excitations and the two-mode squeezing are virtual. They only exist on the level of the noninteracting fields $\hat a$ and $\hat b$ with frequencies $\omega$ and $\Omega$. In the ultrastrong coupling limit, however, these modes {no longer} correspond to any directly measurable quantities{, i.e.}, no correlation between $\hat a$ and $\hat b$ can be detected directly. 
The Hamiltonian that describes measurable quantities can be constructed
with a standard Bogoliubov transformation 
\begin{align}
    \hat H = \omega_- \hat e^\dagger \hat e + \omega_+ \hat f^\dagger \hat f,
\end{align}
where
\begin{align}
\omega_\pm = \sqrt{\frac{1}2\left( \omega^2 + \Omega^2 \pm \sqrt{\left(\omega^2 - \Omega^2\right)^2 + 4  g^2 \omega \Omega}\right)},
\end{align}
are the new frequencies of the physical {(normal)} modes $\hat e$ and $\hat f$, typically dubbed as polaritons in the context of light-matter systems~\cite{BasovAsenjoGarciaSchuckZhuRubio2021}.
Note that the normal frequencies are nontrivial functions of $\omega$ and $\Omega$. In the resonant case, where $\omega = \Omega$, the normal modes are described by 
\begin{align}
 \hat e =  \exp\left[{\frac{\xi_-}{2}\left(\hat c^{\dagger 2} - \hat c^{ 2}\right)}\right] \, \hat c  \, \exp\left[{\frac{\xi_-}{2}\left(\hat c^2 - \hat c^{\dagger 2}\right)}\right], 
\end{align}
and
\begin{align}
\hat f =  \exp\left[{\frac{\xi_+}{2}\left(\hat d^{\dagger 2} - \hat d^{ 2}\right)}\right] \, \hat d \, \exp\left[{\frac{\xi_+}{2}\left(\hat d^2 - \hat d^{\dagger 2}\right)}\right].
\end{align}
Their corresponding frequencies simplify to
\begin{align}
    \omega_\pm = \omega\sqrt{1 \pm \frac{g}{\omega} } = \omega\sqrt{1 \pm \frac{g}{g_c} } = \omega \exp(2 \xi_\pm),
\end{align}
and are related to the number of virtual excitations in modes $\hat c$ and $\hat d$ through the squeezing parameters
\begin{align}
    \langle \hat c^\dagger \hat c \rangle = \sinh^2\xi_- \hspace{0.5cm} \text{and} \hspace{0.5cm}  \langle \hat d^\dagger \hat d \rangle = \sinh^2\xi_+.
\end{align}

{In} terms of the normal modes, the ground state contains no excitations,
\(
    \langle \hat e^\dagger \hat e \rangle = \langle \hat f^\dagger \hat f \rangle = 0,
\)
as it should be for the ground state of the system. At first glance, this virtual two-mode squeezing might seem practically unusable. Converting the virtual excitations into real, physical ones could, in principle, be achieved using techniques such as the dynamical Casimir effect~\cite{nori2011dynamiclcaeff,physics2010007}, for example, by rapidly modulating the light-matter interaction strength $g$. However, as we demonstrate below, this conversion is not strictly necessary as modifications to system properties induced by these correlated virtual excitations---{namely, the frequency shifts}---can themselves be leveraged for applications in quantum sensing (see Fig.~\ref{fig:schematic} for a schematic illustration).


\begin{figure}[htb!]
    \centering
    \includegraphics[width=1\linewidth]{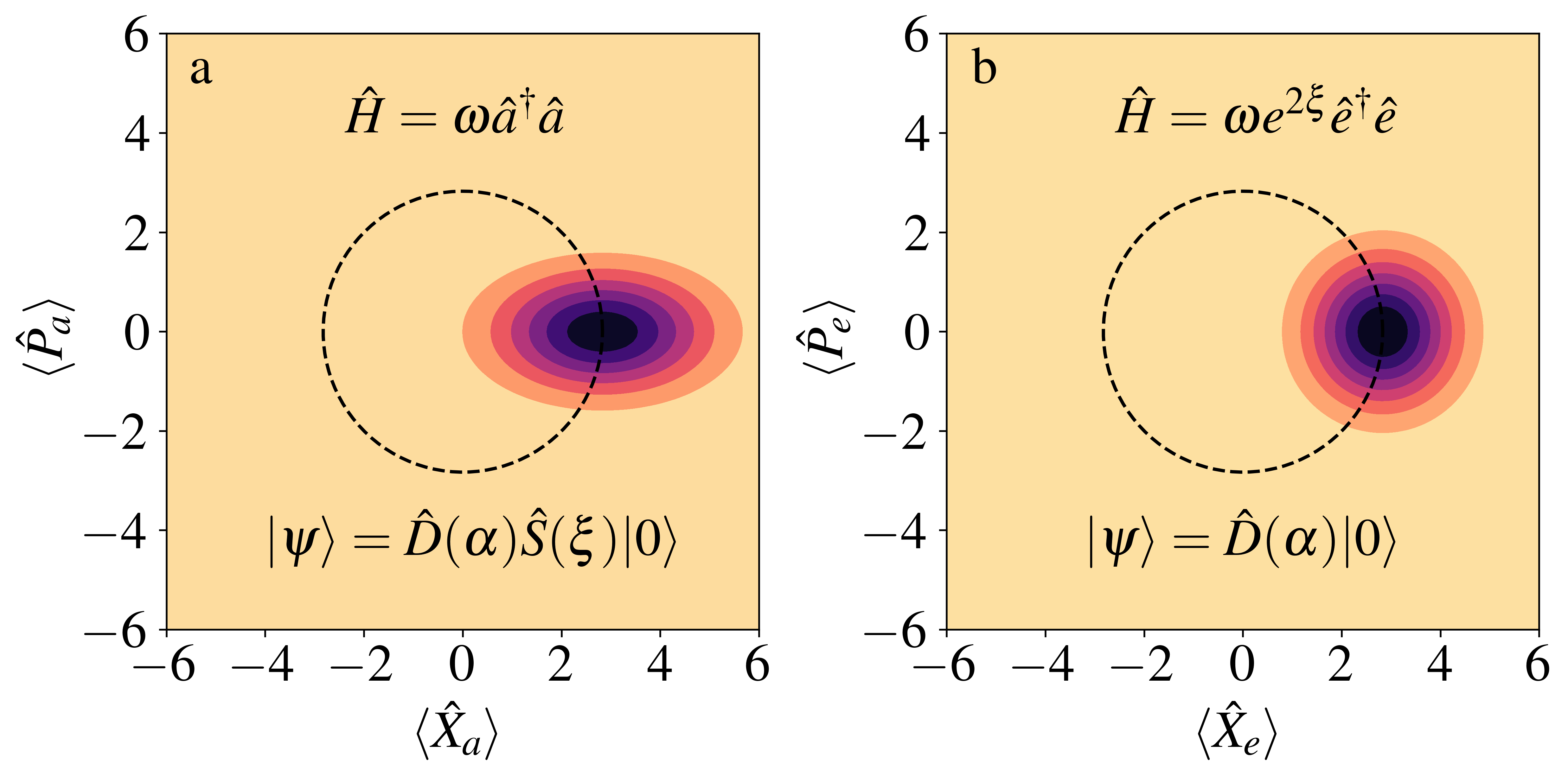}
    \caption{Schematic illustrating the concept. (a) In conventional quantum metrology, explicit squeezing reduces quantum noise along one direction. 
    (b) Alternatively, virtual squeezing, manifested as a modified system frequency, can be harnessed. In this approach, the signal is amplified through a non-trivial dependence of the effective frequency, $\omega e^{2\xi}$, on the unknown parameter $\omega$ through the squeezing parameter $\xi$. For the same level of squeezing, characterized by $\xi$, and simple quadrature measurements, virtual squeezing offers improved sensitivity~\cite{[{See Supplemental Material at }][{ for details on (i) the calculation of quantum Fisher information for a
coherent state in the driven-dissipative ultrastrongly coupled
Dicke model in the large $N$ limit, and (ii) a comprehensive
analysis of quantum metrology based on the quantum Rabi model with
virtual excitations..}]sup1} and exhibits greater robustness against noise and decoherence~\cite{DeLiberato2017virtualphotonosdisspative}.}
    \label{fig:schematic}
\end{figure}

\emph{Quantum Fisher information perspective.}---Given the symmetry between $\omega$ and $\Omega$, we focus on measuring the frequency of the non-interacting radiation mode $\omega$ only.  By either directly measuring or converting the virtual excitations into real ones, their squeezing properties can be harnessed for precision measurements. For example, in critical metrology~\cite{zanardi2007criticaltensors,paris2014LMGcrit,zakrzewski2018QCMlimits,garbe2020criticalmet,Gietka2021adiabaticcritical,gietka2023squezeingsocbec,marti2023CQMfeedback}, the ground state of the system, $|\psi_0\rangle$, near the critical point of a phase transition is used to calculate the quantum Fisher information  with respect to $\omega$
\begin{align}\label{eq:qfi}
    \mathcal{I}_\omega = 4\left(\langle \partial_\omega \psi_0|\partial_\omega \psi_0\rangle  - |\langle \partial_\omega \psi_0|\psi_0\rangle|^2\right),
\end{align}
which determines the sensitivity of estimating the parameter $\omega$ through the quantum Cramér-Rao bound $\Delta \omega \geq 1/\sqrt{\mathcal{I}_\omega}$. For the Dicke model’s ground state, the quantum Fisher information is given by
\begin{align}
    \mathcal{I}_\omega = 2\left(\partial_\omega \xi_-\right)^2 + 2\left(\partial_\omega \xi_+\right)^2,
\end{align}
which reaches its maximum near the critical point of the superradiant phase transition, where
\begin{align}
    \mathcal{I}_\omega \approx 2\left(\partial_\omega \xi_-\right)^2 = \frac{g^2}{8 \omega^4 (1-g/g_c)^2}.
\end{align}
However, leveraging this sensitivity relies on the ability to measure virtual excitations represented by $\hat{a}$ and $\hat{b}$~\cite{solano2019rmp,Falci2024virtualPhot}. 

Here, we demonstrate that transforming virtual particles into measurable ones is not a prerequisite for utilizing them in
precision measurements. Instead, measuring the modified properties of physical systems can already yield significant enhancements over weakly and strongly coupled light-matter systems. In order to measure the properties of hybrid modes, the system must be probed. This can be accomplished by using a laser drive tuned near the resonance frequency of a normal mode. When the modes are spectrally well-separated, focusing on a single mode becomes feasible. Specifically, we concentrate on the lower polariton mode, as its
frequency exhibits a pronounced sensitivity to variations in the bare frequency $\omega$, particularly near the superradiant threshold. The Hamiltonian describing such an isolated normal mode system is given by 
\begin{align}\label{eq:polaritonH}
    \hat H = \omega_- \hat e^\dagger \hat e.
\end{align} 
For simplicity, we now assume that the system is excited into a coherent state $|\alpha\rangle$, satisfying
$\hat{e}|\alpha\rangle=\alpha|\alpha\rangle$, where $|\alpha|^2$ is the number of real (measurable) excitations. The subsequent dynamic is governed by the Schr\"odinger equation using Hamiltonian~\eqref{eq:polaritonH}. Under these assumptions, the state of the system as a function of time is given by
\begin{align}
    |\psi(t)\rangle = \exp(-i \omega_- t \hat e^\dagger \hat e)|\alpha\rangle = |\alpha \exp(-i \omega_- t)\rangle.
\end{align}
Calculating the quantum Fisher information with respect to the parameter of interest $\omega$ yields
\begin{align}
    \mathcal{I}_\omega =  4 t^2 \times |\alpha|^2  \times \left(\partial_\omega \omega_-\right)^2,
\end{align}
which we deliberately separate into three components. The first term, $t^2$, highlights that better
resolution can be achieved by allowing the system to interact with a perturbation for a longer duration. The second term,
$|\alpha|^2$, shows that sensitivity in parameter estimation can be improved by optimizing the initial state. The third term indicates that sensitivity is further enhanced when the frequency is a function of $\omega$ with a large derivative. While the first two terms are commonly exploited to enhance sensitivity~\cite{LLOYD2006QM,lloyd2011advancesQM,smerzi2018rmp,photonicQM2020sciarrino}, the potential of the third term to increase the quantum Fisher information has been so far overlooked. In the weak and strong coupling regimes, frequency shifts exhibit only weak dependence on the field frequency~\cite{helmut2013rmp}. As a result, the derivative of the effective frequency with respect to the bare frequency remains close to unity, limiting the quantum Fisher information to the standard quantum limit
\(
    \mathcal{I}_\omega =  4 t^2 \times |\alpha|^2.
\)
This is not the case in the ultrastrong coupling limit, where the derivative can be extremely large close to the threshold point $g\sim g_c$
\begin{align}
    \partial_\omega \omega_- = 
    \frac{2 -g/\omega}{2  \sqrt{1-\frac{g}{\omega }}}
\end{align}
and the quantum Fisher information becomes
\begin{align}
    \mathcal{I}_\omega =  4 t^2 \times |\alpha|^2  \times \frac{(2-g/\omega )^2}{4   (1-g/\omega )},
\end{align}
which for $g \approx g_c$ can be approximated as
\begin{align}
    \mathcal{I}_\omega \approx 4 t^2 \times |\alpha|^2  \times \frac{1}{4  \left(1-\frac{g}{g_c}\right)}.
\end{align}
The above quantum Fisher information can surpass the standard quantum limit, which can only be achieved with nonclassical resources~\cite{smerzi2009entangHL,smerzi2018rmp}. 
The enhancement originates from the quantum effects of correlated virtual excitations which form the hybrid light-matter modes
\begin{align}\label{eq:virt_sqz}
    \langle  \hat n_c \rangle \equiv \langle \hat c^\dagger \hat c \rangle  = \sinh^2 \xi_- = \frac{\left(1- \sqrt{1-\frac{g}{g_c }}\right)^2}{4 \sqrt{1-\frac{g}{g_c }}}.
\end{align}
{Therefore, the quantum Fisher information can be expressed with the virtual excitations $\langle \hat n_c \rangle$ {or, equivalently, the virtual squeezing parameter $\xi_-$} as
\begin{align}
     \mathcal{I}_\omega \approx 16 t^2 \times |\alpha|^2  \times \langle \hat n_c \rangle^2 =   t^2 \times |\alpha|^2 \times  {\exp(-4\xi_-)},
\end{align}
exhibiting Heisenberg scaling with respect to the number of virtual excitations~\cite{demko2012elusiveHL}.} {Remarkably, suppressing noise using the same squeezing parameter $\xi_-$—which is equivalent to extracting the squeezed virtual excitations from the ground state and displacing it by $\alpha$ (see Fig.~\ref{fig:schematic}a)—yields  
\begin{align} 
\mathcal{I}_\omega  \approx 16 t^2 \times |\alpha|^2 \times \langle \hat n_c\rangle = 4 t^2 \times |\alpha|^2 \times \exp(-2 \xi_- ), \end{align}
leading to quantum Fisher information {that scales only linearly} with the number of virtual excitations. In other words, for a fixed squeezing parameter $\xi_-$, modifying the oscillator frequency---thereby keeping the squeezing excitations virtual---enhances the quantum Fisher information more effectively than reducing quantum noise via real squeezing (see also Supplemental Material~\cite{sup1} for a more detailed comparison). This indicates that retaining squeezing in its virtual form can be
more advantageous than extracting and directly converting the
virtual excitations into real ones.}

However, this enhancement requires operating very close to the critical point, in accordance with the principles of critical metrology~\cite{zakrzewski2018QCMlimits}. Note that the above analysis assumes an isolated system, coherent evolution, preloaded excitations, and neglects the measurement process. A more realistic experimental approach is discussed in the next section.

 
\emph{Signal-to-noise ratio.}---The quantum Fisher information establishes the fundamental precision bound for parameter estimation, optimized over all measurements~\cite{geometry1994caves}, even the most abstract ones. In essence, {it} 
is the maximum signal-to-noise ratio achievable for a given system~\cite{górecki2024interplaytimeenergybosonic}
\begin{align}
    \mathcal{I}_\omega = \max_{\hat O} S(\omega), \,\,\,\,\, \mathrm{where} \,\,\,\,\, S(\omega) =\frac{|\partial_\omega\langle \hat O \rangle|^2}{\Delta^2 \hat O},
\end{align}
with $\hat O$ being a measured observable. Taking into account practical considerations, one cannot perform arbitrary measurements on the system. In particular, direct measurements on virtual excitations are not feasible. In the following case, however, we always deal with a Gaussian system {and Gaussian states}. Hence, all the information about the parameters of the system is contained in the quadrature operators~\cite{[{See Supplemental Material at }][{ for details on (i) the calculation of quantum Fisher information for a
coherent state in the driven-dissipative ultrastrongly coupled
Dicke model in the large $N$ limit, and (ii) a comprehensive
analysis of quantum metrology based on the quantum Rabi model with
virtual excitations.}]sup1}. {Therefore, homodyne and heterodyne detection schemes constitute
optimal measurements ({saturating the quantum Cram\'er-Rao bound}) in this scenario, for which the quantum
Fisher information equals the signal-to-noise ratio~\cite{gaussian2019}}.

Assuming that the two frequencies $\omega_-$ and $\omega_+$ are well separated, we can {consider} 
only one mode of the system. We focus again on the lower normal mode described by $\hat e$. Therefore, let us calculate the signal-to-noise ratio for the measurement of the $\hat X = \left(\hat e + \hat e^\dagger\right)/2$ quadrature for which we also assume that the system is open and loses excitations---a typical situation in cavity quantum electrodynamics experiments~\cite{esslinger2010dickemodel,esslinger2012rotontype}. In the frame rotating with the frequency of the pumping laser (assumed to be in a coherent state) the Hamiltonian of the system is given by a driven harmonic oscillator,
\begin{align}
    \hat H = \delta \hat e^\dagger \hat e + \eta\left(\hat e + \hat e^\dagger \right),
\end{align}
where $\delta \equiv \omega_p - \omega_-$ is the pump-polariton detuning and  $\eta$ is the pump strength. Using the input-output relations---consistent with those considered in Refs~\cite{Ciuti2006inputoutputUSC, Bamba2012,
Bamba2013, DeLiberato2014}, ensuring the conservation of the total number of polaritons rather than photons---it is straightforward to calculate the quadrature properties of the transmission for the driven-dissipative system~\cite{thompson2023ultranarrow}
\begin{align}
    \langle \hat X \rangle = \sqrt{\kappa}\,|\alpha|\exp(- i \varphi) \hspace{0.5cm} \mathrm{and} \hspace{0.5cm} \Delta^2 \hat X = \frac{1}{4 t},
\end{align}
where $|\alpha| = 2 \sqrt{\eta^2/(\kappa^2 +4 \delta^2 )}$ {is the amplitude of the measurable photon field} which depends on the unknown parameter $\omega$ through $\delta(\omega)$, $\kappa$ is the cavity loss rate, $\varphi = \arctan[{\kappa}/{2\delta}]$ is the phase shift between the drive and the system response, and $t$ is the time of the measurement. In the following, we focus on the amplitude measurement~\cite{sup1}.

The signal-to-noise ratio of the quadrature amplitude after a measurement time $t$ becomes~\cite{sup1}
\begin{align}
    S(\omega) = 4 \kappa \left(\partial_\omega|\alpha|\right)^2= \frac{256 \kappa t \delta ^2 \eta ^2 (g/\omega-2  )^2}{  \left(4 \delta ^2+\kappa ^2\right)^3 (1 -g/\omega)},
\end{align}
which can be rewritten including the number of virtual excitations~\eqref{eq:virt_sqz} close to the threshold point as
\begin{align}\label{eq:snr_scaling}
    S(\omega) &\approx \frac{64 \kappa  \delta ^2  }{  \left(4 \delta ^2+\kappa ^2\right)^2 }|\alpha|^2\exp(4\xi_-) t \approx  \frac{2^{10} \kappa t \delta ^2 |\alpha|^2 }{  \left(4 \delta ^2+\kappa ^2\right)^2 }|\alpha|^2\langle \hat n_c\rangle^2 t.
\end{align}
The signal-to-noise ratio from the above equation is presented in Fig.~\ref{fig:amp}. Although the largest amplitude requires the resonance condition $\delta=0$, {this is not the case for the signal as at resonance the amplitude does not depend on detuning, $S(\delta = 0) = 0$.} 
For small couplings, the signal grows slowly [see Fig.~\ref{fig:amp}(a)], and for large couplings the signal grows dramatically near to the threshold point [see Fig.~\ref{fig:amp}(b)] reflecting macroscopic occupation of virtual excitations. 

{{In contrast, reducing the noise using the same squeezing parameter
$\xi_-$---which corresponds to extracting the squeezed virtual
excitations from the ground state---leads to
\begin{align}
    \Delta^2 \hat X = \frac{\exp(2\xi_-)}{4 t} = \frac{\sqrt{1-g/g_c}}{4  t} \approx \frac{1}{16 \langle \hat n_c \rangle t}
\end{align}
and thus results in the signal-to-noise ratio that scales linearly
with the number of virtual excitations, contrary to the
quadratic scaling shown in Eq.~\eqref{eq:snr_scaling}.}}

\begin{figure}[htb!]
    \centering
    \includegraphics[width=1\linewidth]{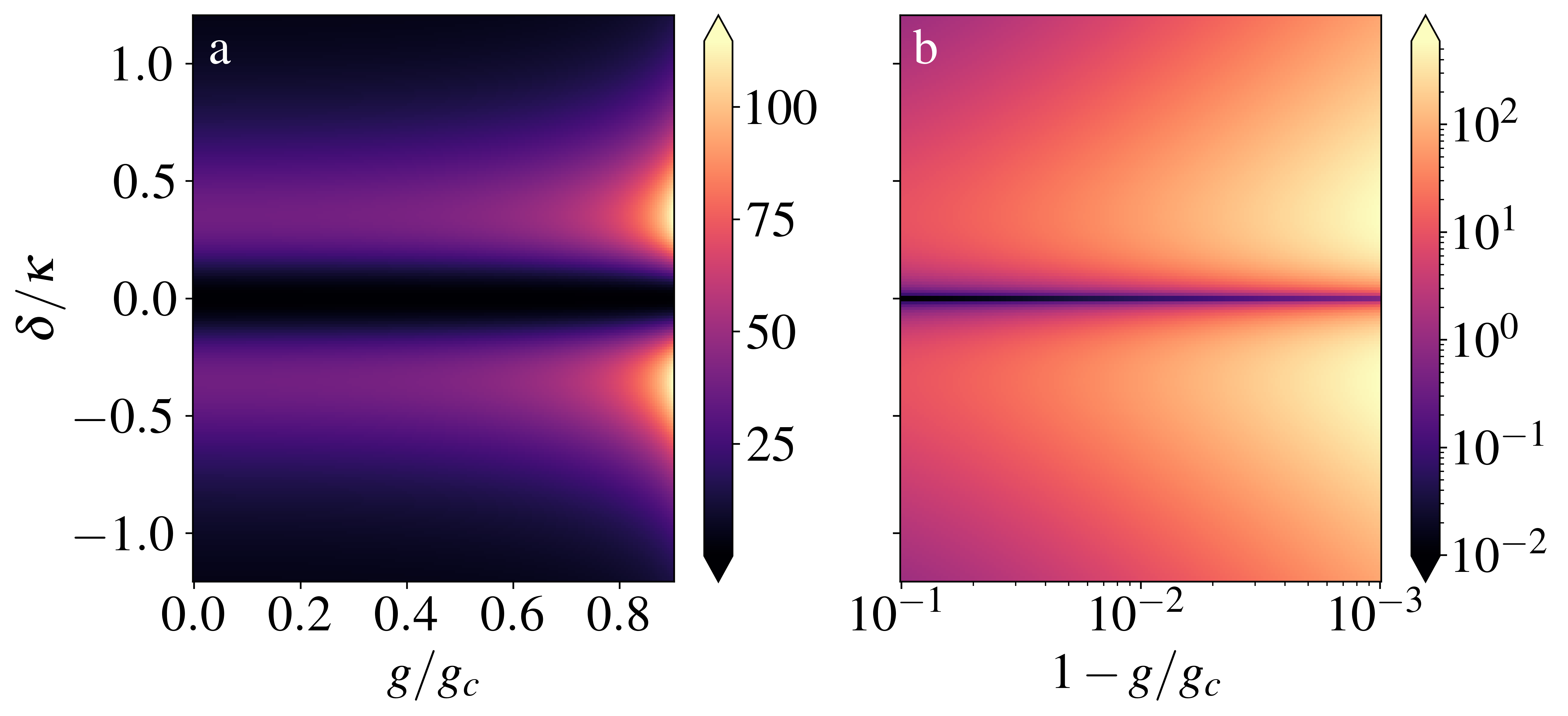}
    \caption{
    The signal-to-noise ratio for amplitude measurements normalized to time $t$ as a function of detuning $\delta/\kappa$ and the coupling strength $g/g_c$. At the resonance ($\delta/\kappa = 0$), the signal does not depend on the unknown parameters, so a slightly off-resonant drive is required for amplitude measurement. (a) At the optimal points, the enhancement is related to the derivative of the effective frequency with respect to the bare frequency. (b) A log-scale zoom near the critical point, where the derivative is extremely large, illustrating the macroscopic occupation of virtual excitations. In the numerical simulation we set $\kappa/\omega =1$ and $\eta/\omega = 1$.}
    \label{fig:amp}
\end{figure}

 
\emph{Limitations.}---The calculations presented so far assume an infinite number of atoms, where the collective spin is approximated as a harmonic oscillator. This approximation, enabled by the Holstein-Primakoff transformation, is typically well-justified in experiments with large atomic ensembles~\cite{Mivehvar02012021}, but it has limitations. In the thermodynamic limit, the spectrum is fully harmonic. Reducing the number of atoms introduces anharmonicity, which can give rise to phenomena such as photon blockade~\cite{kimble2005photonblockade}, where increasing the driving power no longer leads to a proportional increase in output due to saturation and nonlinear effects. As a result, the magnitude of the coherent field at the system's output is inherently limited. {Furthermore, in finite systems, the energy gap cannot fully close, thereby constraining the maximum achievable derivative of the energy gap (frequency) {and the virtual squeezing parameter (see Supplemental Material~\cite{sup1}).}}

Additionally, our calculations neglected the {diamagnetic $A^2$-term} in the Hamiltonian~\eqref{eq:HDickeapprox}. Including this term results in the Hopfield model~\cite{hopfield1958hopfieldmodel}, where the Bogoliubov transformation shows that the cavity frequency increases as $\omega \rightarrow \tilde{\omega} > \omega$. This shift leads to an increase in the critical coupling strength, such that $g_c \rightarrow \tilde{g}_c = \sqrt{\tilde{\omega} \Omega} > \sqrt{\omega \Omega}$. However, as long as there is no upper limit on the coupling strength $g$, the diamagnetic term does not fundamentally prohibit the use of virtual excitations in quantum metrology. {Note that the diamagnetic term is not inherent to light-matter systems. For instance, it does not appear in optomechanical systems~\cite{huard2018effectiveUSC,oriol2022mechanicalsqueezincUSC,uros2023dissipativephasetrans,UrosDelic2024OPtomochenicalUSC}}.

Another essential consideration is the precise knowledge of system parameters, apart from the unknown quantity being measured. Effective use of the signal-to-noise ratio (via the error propagation formula) requires exquisite control over one of the subsystems and the coupling strength~\cite{mihailescu2024uncertainquantumcriticalmetrology}. This reveals a trade-off: leveraging hybrid modes in precision metrology requires near-perfect control and knowledge of one mode to extract maximal information about the other.

{We also emphasize that while critical slowing down is a common feature near phase transitions, it does not universally limit the practical implementation of our scheme. In particular, for large coupling strengths \( g \gg g_c \), the upper polariton mode becomes the relevant excitation, and the associated energy gap increases with \( g \), avoiding divergent relaxation times. Moreover, in some strongly correlated platforms, such as polaritons in condensed matter~\cite{bastard2005vacuumproperties}, rapid thermalization can occur despite a vanishing gap, due to many-body effects. These scenarios demonstrate that enhanced sensitivity near criticality can, in certain regimes, coexist with favorable dynamical properties, ensuring practical applicability of our approach.}

Finally, it is {{important to emphasize}} that resonant conditions are not strictly required to exploit virtual excitations and similar results can be obtained in the dispersive regime. For instance, in the quantum Rabi model, {{large}} detunings ($\omega \ll \Omega$) are necessary to maximize virtual excitations in the normal phase~\cite{plenio2016qptRM}. In this regime, the mode with the lower frequency exhibits virtual squeezing~\cite{[{See Supplemental Material at }][{ for details on (i) the calculation of quantum Fisher information for a
coherent state in the driven-dissipative ultrastrongly coupled
Dicke model in the large $N$ limit, and (ii) a comprehensive
analysis of quantum metrology based on the quantum Rabi model with
virtual excitations..}]sup1}. {{Likewise}}, in the Dicke model, the ratio $\omega / \Omega$ {{dictates}} which mode accumulates more virtual excitations~\cite{emarybrandes2003dickechaosphase,gietka2021invertedoscillator}. In the extreme limit of $\omega / \Omega \rightarrow 0$, only the cavity mode becomes significantly populated with virtual excitations. {This {{also implies}} that $\Omega$ can be treated as a control parameter when $g$ is fixed and well known, in the case of $\omega$ being the to-be-measured parameter.}


\emph{Conclusions.}---In this work, we demonstrate that {{virtual excitations}—a hallmark of the ultrastrong coupling regime in light-matter systems—can be harnessed} to enhance quantum precision measurements. Using the Dicke model as a testbed, we show that these {virtual processes reshape the system’s eigenstructure and modify the curvature of the Hamiltonian’s parameter manifold}, thereby boosting the {{quantum Fisher information}}~\cite{zanardi2007criticaltensors,Zanardi2007Informationgeometry} {and increasing the information content of the quantum state}.

{In contrast to conventional squeezing, which requires precise external control, our approach leverages {intrinsic squeezing} that naturally arises in ultrastrongly coupled systems.} The interaction between an unknown subsystem and a well-characterized one creates a {{hybrid mode}} whose {{renormalized frequency}—shifted due to virtual excitations—acts as a robust and experimentally accessible signal carrier}~\cite{RevModPhys.78.1297}. {This frequency can then be measured using established quantum metrology techniques}~\cite{hotter2024combining}.

{More broadly, this strategy provides a powerful sensing framework. By coupling an unknown quantum system to a controlled probe, the emergent hybrid mode encodes information in a way that is both sensitive and resilient.} {This is especially advantageous for macroscopic quantum systems, where traditional squeezing protocols face technical limitations.} {Our results suggest that ultrastrong coupling offers not just an alternative to engineered squeezing, but a {complementary path} to quantum-enhanced sensitivity.} {Importantly, our proposal is experimentally viable, with proof-of-concept implementations already within reach in platforms such as} cavity QED~\cite{esslinger2010dickemodel,esslinger2012rotontype,hemmerich2015dickedynamic,lev2017phasetransitionpolariton}, circuit QED
\cite{Yoshihara2016,Gu2017}, optomechanics~\cite{uros2023dissipativephasetrans,delic2024USCoptomechanics}, {and trapped ions}~\cite{Ustinow2017USC_QRM,rey2018dickeion,bollinger2021sensingdicke,duan2021qrmsingleion}.

\acknowledgments 
The authors would like to acknowledge Simon Hertlein, Farokh Mivehvar, and Helmut Ritsch for discussions. A.M. was supported by the Polish National Science Centre (NCN) under the Maestro Grant No. DEC-2019/34/A/ST2/00081. C.H. was supported by the Carlsberg Foundation through the “Semper Ardens” Research Project QCooL. Simulations were performed using the open-source \textsc{QuantumOptics.jl}~\cite{kramer2018quantumoptics} framework in \textsc{Julia}.

%



\renewcommand\theequation{S\arabic{equation}}
\setcounter{equation}{0}

\renewcommand\thefigure{S\arabic{figure}}
\setcounter{figure}{0}

\renewcommand\thesection{S\arabic{section}}
\setcounter{section}{0}

\maketitle

\onecolumngrid


\section{S1. Quantum Fisher information for a coherent state of driven-dissipative ultrastrongly coupled Dicke model in the limit of number of atoms \(N\) going to $\infty$}
\begin{figure}[htb!]
    \centering
    \includegraphics[width=1\linewidth]{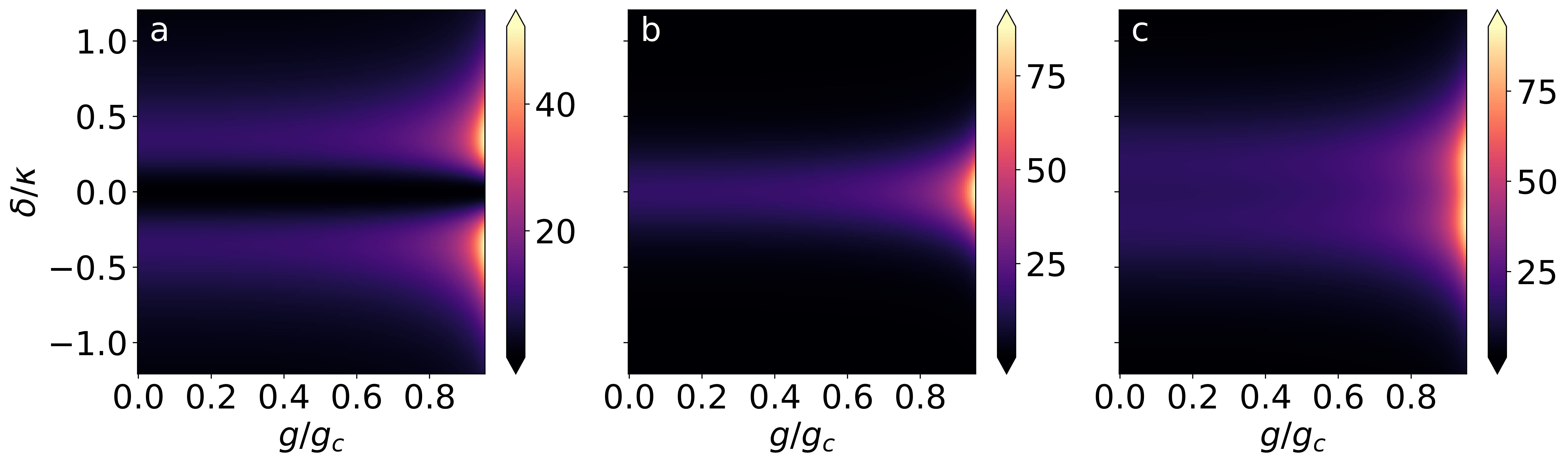}
    \caption{
    The signal-to-noise ratio for (a) amplitude measurements (described in the main text), (b) phase measurements, and (c) combined amplitude and phase measurements (quantum Fisher information) as a function of $g/g_c$ and $\delta/\kappa$. In the numerical simulations, we set $\kappa/\omega =1$ and $\eta/\omega = 1$. Note that in the context of cavity QED, one could also perform measurements on the reflected light.}
    \label{fig:compAnadP}
\end{figure}

\noindent The quantum Fisher information can also be calculated using the coherent state output (transmission) as
\begin{align}
    \mathcal{I}_\omega = 4 \left(\partial_\omega \alpha\right)^2 =  4  \left(\partial_\omega A\right)^2 + 4  A^2 \left(\partial_\omega \varphi\right)^2,
\end{align}
where $A$ and $\varphi$ is the amplitude and phase of $\alpha \equiv A\exp(i\varphi)$, respectively. For a driven-dissipative system in the frame rotating with the driving field frequency and using input-output relations
, it is straightforward to show
\begin{align}
    \alpha = 2  \sqrt{\frac{\kappa\eta^2}{\kappa^2 +4 \delta^2 }} \, {\exp\left[ -i \arctan\left(\frac{\kappa}{2\delta}\right)\right]}
\end{align}
such that the number of photons per unit of time at the output is
\begin{align}
    \langle \hat n \rangle/t  =\frac{4 \kappa \eta^2}{\kappa^2 + 4 \delta^2}.
\end{align}
The quantum Fisher information after time $t$ becomes
\begin{align}
    \mathcal{I}_\omega = \frac{4 t  \eta^2 \kappa }{ 4 \delta ^2+\kappa ^2} \times \frac{16 \delta ^2  (g-2 \omega )^2}{\omega  \left(4 \delta ^2+\kappa ^2\right)^2 (\omega -g)} + \frac{4 t  \eta^2 \kappa }{ 4 \delta ^2+\kappa ^2} \times \frac{\kappa ^2 (g-2 \omega )^2}{\omega  \left(4 \delta ^2+\kappa ^2\right)^2 (\omega -g)},
\end{align}
where the first term corresponds to the information extractable from the amplitude (see the main text), while the second term corresponds to the information extractable from the phase shift. A comparison of the information extractable from the amplitude and phase is shown in Fig.~\ref{fig:compAnadP}. 

{For $g=0$, we get the known result for quantum Fisher information:

\begin{align}
    \mathcal{I}_\omega = \frac{256\kappa t \delta ^2 \eta^2}{\left(4 \delta ^2+\kappa ^2\right)^3 } + \frac{16 t  \eta^2 \kappa ^3 }{ \left(4 \delta ^2+\kappa ^2\right)^3 }.
\end{align}
The enhancement deriving from a non-zero $g$ can be linked to the presence of virtual excitations as described in the main text and in the following sections.}

\section{S2. A simplified example: The quantum Rabi model with virtual excitations}
\noindent In the main text, we focus on the Dicke model, which is an extension of the quantum Rabi model including many two-level systems. Here, we show how one could use virtual photons in the quantum Rabi model to enhance the precision of a cavity frequency measurement. 
The Hamiltonian of the quantum Rabi model reads
\begin{align}
    \hat H = \omega \hat a^\dagger \hat a + \frac{\Omega}{2}\hat \sigma_z + \frac{g}{2}\left(\hat a + \hat a^\dagger\right) \hat \sigma_x,
\end{align}
where the Pauli matrices $\hat \sigma_i$ ($i=x,y,z$) represent the two-level system degree of freedom. A crucial step here is to operate in the regime where the energy gap could be almost closed, i.e.\ we require $\Omega \gg \omega$. Essentially, the smaller the energy gap is, the more virtual photons the system can support. {In order to enhance the measurement precision of $\omega$, the energy gap needs to depend strongly on $\omega$. In {Fig.~\ref{fig:qrm},} we plot the derivative of the energy gap between the first excited state and the ground state $\Delta E = E_1 - E_0$ with respect to $\omega$. Close to the critical point ($g \lesssim g_c$) $\partial_\omega \Delta E$ is larger for larger values of $\Omega / \omega$.} 
Note that near the superradiant threshold, $g\approx g_c \gg \omega$ is no longer comparable with the transition frequency $\omega$. Therefore, this regime is called the deep ultrastrong coupling regime. In the regime of $\Omega\gg\omega$, one can apply the Schrieffer-Wolff transformation and obtain {(exact in the limit of $\Omega/\omega \rightarrow\infty$ )}
\begin{align}
    \hat H = \omega \hat a^\dagger \hat a -\frac{g^2}{4\Omega}\left(\hat a + \hat a^\dagger\right)^2,
\end{align}
which is a squeezing Hamiltonian, whose ground state is the squeezed vacuum
\begin{align}
    |\psi_0\rangle = \hat S(\xi)|0\rangle,
\end{align}
where $\hat S(\xi)$ is the squeeze operator with the squeezing parameter $\xi = \frac{1}4\log\left(1-g^2/g_c^2\right)$ and $g_c = \sqrt{\omega \Omega}$, which contains 
\begin{align}
    \langle \hat n \rangle  \equiv \langle \hat a^\dagger \hat a \rangle =\sinh^2 \xi \xrightarrow[]{g\rightarrow g_c} \frac{1}{4}\frac{1}{\sqrt{1-\frac{g^2}{g_c^2}}}
\end{align}
virtual photons. In the picture of measurable quantities, one has to diagonalize the Hamiltonian
\begin{align}
    \hat H = \omega\sqrt{1-g^2/g_c^2}\,\hat c^\dagger \hat c,
\end{align}
where $\hat c = \hat a \cos \xi + \hat a^\dagger \sin \xi$ represents a cavity mode of frequency $\omega\sqrt{1-g^2/g_c^2}\equiv \omega \exp(2\xi)$, measurable in experiments. For the ground state, we have $\langle \hat c^\dagger \hat c \rangle =0$. Note that a similar effective Hamiltonian can also be used to describe the Dicke model far from the resonance. For simplicity, we assume that there is no decoherence and no dissipation, and an arbitrary measurement can be performed. We consider three possible scenarios, which are compared in Fig.~\ref{fig:comparison}.

{\begin{figure}[htb!]
    \centering
    \includegraphics[width=0.4\linewidth]{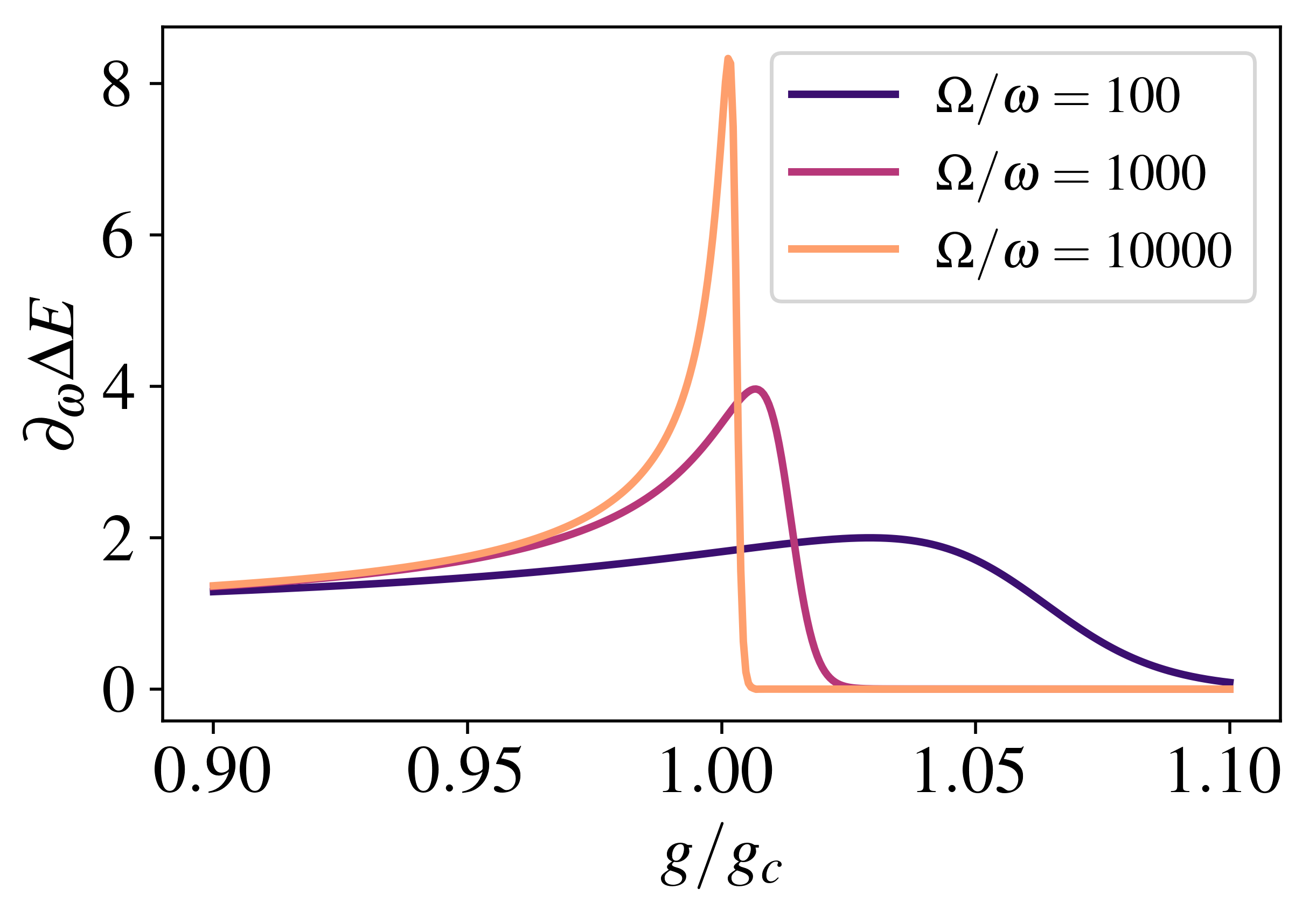}
    \caption{Derivative of the energy gap with respect to $\omega$ as a function of $g/g_c$ for three cases: $\Omega/\omega =100$ (violet), $\Omega/\omega=1000$ (purple), and $\Omega/\omega=10000$ (orange). The larger the ratio $\Omega/\omega$, the tighter the energy gap can get, and therefore the more virtual photons the system can support. Note that this is the energy gap between the ground state and the first excited state $\Delta E = E_1 -E_0$. In the thermodynamic limit defined as $\Omega/\omega \rightarrow \infty$, the spectrum is fully harmonic (linear), however, this is not the case for a finite $\Omega/\omega$ putting a limitation on how strongly the system can be excited with a coherent drive. 
    }
    \label{fig:qrm}
\end{figure}}

\newpage
\subsection{S2.1 Extracting the squeezed vacuum}
\noindent First, we consider a scenario where one could extract the squeezed vacuum state from the ground state and use it to estimate the frequency of the harmonic oscillator. In such a case the signal-to-noise ratio can be calculated using the quadrature operators for mode $\hat a$. Since for squeezed vacuum the first moment of $\hat X = \left(\hat a+ \hat a^\dagger \right)/2$ is zero, we must rely on the second moment,
\begin{align}
    \langle \hat X^2 \rangle = \frac{1}{4}\exp(- 2 \xi) = \frac{1}{4}\frac{1}{\sqrt{1-\frac{g^2}{\omega \Omega}}},
\end{align}
and the fourth moment,
\begin{align}
    \langle \hat X^4 \rangle = 3     \langle \hat X^2 \rangle^2 = \frac{3}{16}\exp(-4 \xi) = \frac{3}{16}\frac{1}{1 - \frac{g^2}{ \omega  \Omega}},
\end{align}
to calculate the signal-to-noise ratio. By calculating the derivative with respect to $\omega$ of the second moment, we obtain the following signal-to-noise ratio:
\begin{align}
    S(\omega) = \frac{|\partial_\omega \langle \hat X^2\rangle|^2}{\langle \hat X^4\rangle - \langle \hat X^2 \rangle^2}= \frac{1}{8 \omega^2 \left(1 - \frac{g ^2}{g_c^2} \right)^2} \frac{g ^4}{g_c^4}. 
\end{align}
However, this signal-to-noise ratio assumes that one can either extract virtual excitations from the ground state or perform a measurement in the basis of the virtual excitations.

The signal-to-noise ratio could be further increased by allowing the squeezed vacuum to evolve in the noninteracting harmonic oscillator for time~$t$. In this case, the second moment becomes
\begin{align}
    \langle \hat X^2 \rangle = \frac{1}{4}\left[ \exp(-2\xi)\cos^2 (\omega t) + \exp(2\xi)\sin^2(\omega t)\right],
\end{align}
and the signal-to-noise ratio can be calculated to be
\begin{align}\label{eq:case1}
    S(\omega) = \frac{1}{8} \left[\frac{4 \left(g^2 t\sin (2  \omega t)-\Omega \right)}{g^2 \cos (2  \omega t)-g^2+2 \omega  \Omega }+\frac{g^2-2 \omega  \Omega }{g^2 \omega -\omega ^2 \Omega }\right]^2.
\end{align}
%
Note that, in certain situations, measuring the second moment rather than the first may be significantly more time-consuming.


\subsection{S2.2 Extracting the squeezed vacuum and displacing it}
\noindent After extracting the virtual excitations, one could displace them and utilize the displaced squeezed vacuum for measurement. Consequently, the final state is 
\begin{align}
    |\psi\rangle = \hat D(\alpha)\hat S(\xi)|0\rangle.
\end{align}
In this case, the signal-to-noise ratio can be calculated using the first moment 
\begin{align}
    \langle \hat X \rangle = |\alpha|\cos \omega t,
\end{align}
and the variance
\begin{align}
    \langle \hat X^2 \rangle - \langle \hat X \rangle^2 = \frac{1}{4}\exp(-2 \xi) \cos^2( \omega t) +\frac{1}{4}\exp(2 \xi) \sin^2( \omega t). 
\end{align}
Combining all the elements, we arrive at the following limitation for the signal-to-noise ratio:
\begin{align}\label{eq:case_B}
    S(\omega) = \frac{|\partial_\omega \langle \hat X\rangle|^2}{\langle \hat X^2\rangle -\langle \hat X\rangle ^2} 
    = \frac{8 |\alpha| ^2 t^2  \sqrt{1-\frac{g^2}{g_c^2 }} \sin ^2( \omega  t)}{\frac{g^2}{g_c^2 } \cos (2  \omega t)-\frac{g^2}{g_c^2 }+2 }\leq \frac{4|\alpha|^2 t^2 }{\sqrt{1-\frac{g^2}{g_c^2}}} \xrightarrow[]{g\rightarrow g_c} {16|\alpha|^2 t^2 }\langle \hat n \rangle.
\end{align}


\subsection{S2.3 Measuring the new normal mode}
\noindent Alternatively, instead of extracting the virtual photons from the ground state, one could leverage the information about the unknown parameter through the altered properties of the system as outlined in the main text. In this case, the signal-to-noise ratio is calculated for the normal mode $\hat c$ and its quadrature $\hat X_c$. For the first moment, we obtain
\begin{align}
    \langle \hat X_c \rangle = |\alpha| \cos \left(\omega e^{2\xi} t \right),
\end{align}
and for the variance, we obtain
\begin{align}
    \langle \hat X_c^2 \rangle - \langle \hat X_c \rangle^2 = \frac{1}{4}.
\end{align}
By combining all the terms and calculating the derivative with respect to the unknown parameter $\omega$, we find the signal-to-noise ratio is constrained as 
\begin{align}\label{eq:caseC}
    S(\omega) = \frac{ |\alpha|^2 t^2 \sin^2  \omega e^{2\xi} t}{1-\frac{g^2}{g_c^2}} \left(2-\frac{g^2}{g_c^2}\right)^2
    \leq \frac{ |\alpha|^2 t^2 }{1-\frac{g^2}{g_c^2}} \left(2-\frac{g^2}{g_c^2}\right)^2 \xrightarrow[]{g\rightarrow g_c}\frac{ |\alpha|^2 t^2 }{1-\frac{g^2}{g_c^2}} =  16|\alpha|^2 t^2 \langle \hat n \rangle^2.
\end{align}
In the final step, we explicitly incorporated the number of virtual excitations into the analysis. {Note the different scaling with respect to virtual excitations in Eq.~\eqref{eq:case_B} and Eq.~\eqref{eq:caseC}.}

\begin{figure}[htb!]
    \centering
    \includegraphics[width=1\linewidth]{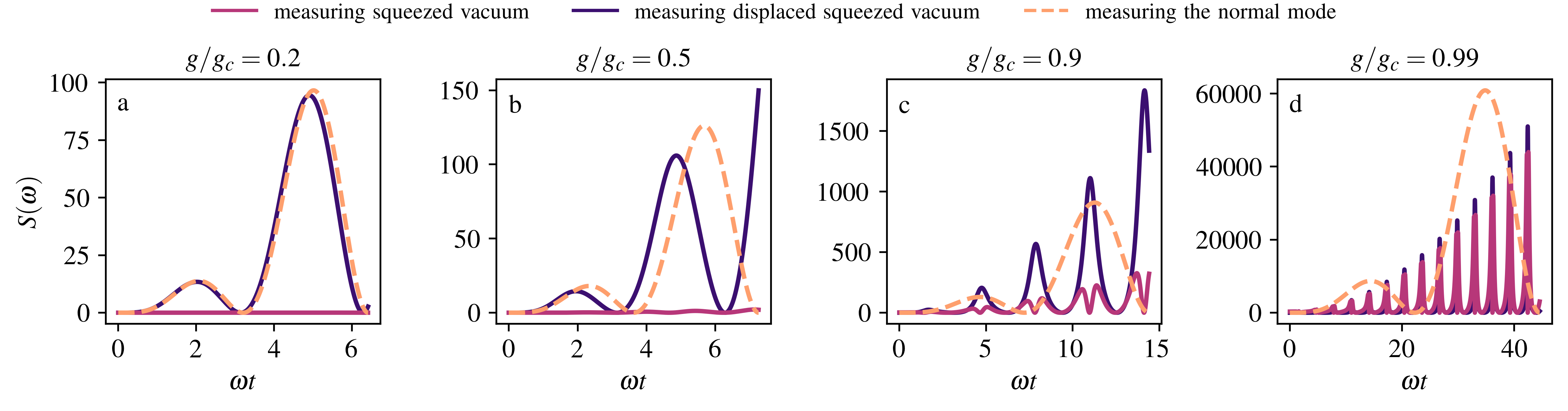}
    \caption{Comparison of the signal-to-noise ratio for the three approaches to precision measurements. Note that since we have Gaussian states, for the optimal quadrature angle {the signal-to-noise ratio becomes the quantum Fisher information which would correspond to the envelope of the curves.}  
    The purple curves represent the strategy of extracting virtual photons and allowing them to evolve within the harmonic oscillator [Eq.~\eqref{eq:case1}]. The violet curves correspond to extracting virtual excitations, displacing them, and then evolving in the harmonic oscillator [Eq.~\eqref{eq:case_B}]. The orange curves illustrate measurements performed on the effective, measurable normal mode without extracting the virtual photons [Eq.~\eqref{eq:caseC}]. When ultimate control is achievable, retaining the virtual excitations within the virtual realm proves advantageous. For $g\ll g_c$ [as seen in (a) for $g=0.2 g_c$], there is no significant enhancement from virtual excitations due to their negligible presence in the system. As the ultrastrong coupling regime emerges (b, $g=0.5 g_c$), virtual excitations begin to appear, contributing to an increased signal-to-noise ratio. Approaching the threshold point (c, $g=0.9 g_c$), it becomes advantageous to extract the virtual excitations from the ground state. However, even closer to the threshold point (d, $g=0.99 g_c$), where correlated virtual excitations are substantially present, retaining them in the virtual domain can be more beneficial for quantum metrology. In the simulations, we set the displacement parameter $\alpha=1$ and assume $\omega/\Omega\rightarrow 0$.}
    \label{fig:comparison}
\end{figure}

\subsection{S2.4 Comparison}
\noindent The comparison between all these cases is presented in Fig.~\ref{fig:comparison}. Interestingly, if the coupling $g$ can be arbitrarily controlled, performing measurements on the normal modes of the coupled system proves to be more advantageous than extracting the virtual excitations. For small couplings $g/g_c$, as shown in Fig.~\ref{fig:comparison}(a) and Fig.~\ref{fig:comparison}(b), the number of virtual excitations is negligible. Consequently, measuring them results in an insignificant signal-to-noise ratio (purple curve). In scenarios where a displacement is applied (violet and orange curves), the signal is predominantly attributed to the displacement. 
As the system approaches the critical point, as illustrated in Fig.~\ref{fig:comparison}(c) for $g/g_c=0.9$, the number of virtual excitations begins to grow, leading to an increased signal-to-noise ratio when they are measured (purple curve). However, when the coupling approaches the critical point of the superradiant phase transition $g/g_c=0.99$ [presented in Fig.~\ref{fig:comparison}(d)], the optimal strategy shifts to measuring the effective mode without extracting the virtual excitations. This approach can be further enhanced by increasing the displacement $|\alpha|$ and applying quantum noise squeezing. 

{\subsection{S2.4 Synergy of virtual and real squeezing}
Finally, we note that the discussed system with a modified
(squeezed) frequency can also be driven by squeezed light,
characterized by a squeezing parameter $\xi_r$. In such a case,
the first moment takes the form:
\begin{align}
    \langle \hat X_c \rangle = |\alpha| \cos \left(\omega e^{2\xi} t \right),
\end{align}
while the variance is given by
\begin{align}
    \langle \hat X_c^2 \rangle - \langle \hat X_c \rangle^2 \geq \frac{1}{4} \exp(2\xi_r).
\end{align}
Combining these results, the signal-to-noise ratio in the limit \(g \rightarrow g_c\) becomes
\begin{align}
    S(\omega) \leq \frac{ |\alpha|^2 t^2 }{1 - \frac{g^2}{g_c^2}} \exp(-2\xi_r) = 16 |\alpha|^2 t^2 \langle \hat n \rangle^2 \exp(2\xi_r) = |\alpha|^2 t^2 \exp(-4\xi) \exp(-2\xi_r)= |\alpha|^2 t^2 \exp(-4\xi - 2 \xi_r) ,
\end{align}
where $\xi$ denotes the virtual squeezing parameter (associated
with the frequency shift), and $\xi_r$ quantifies the real
squeezing of the light injected into the oscillator. This
expression clearly demonstrates that virtual squeezing is more
effective in enhancing the signal-to-noise ratio, and that virtual
and real squeezing can, in principle, be combined to achieve
compounded metrological advantages.

This result highlights a crucial distinction between virtual and
real squeezing in terms of their metrological benefits. Virtual
squeezing boosts the signal, resulting in an exponential
improvement in sensitivity. In contrast, real squeezing only
reduces the noise level, leaving the signal unaffected. Crucially, virtual squeezing naturally arises from the system's
dynamics near criticality, without requiring nonclassical input
states, making it inherently more robust against decoherence and
losses. The exponential difference in scaling with respect to
$\xi$ and $\xi_r$ suggests that metrological protocols operating
near a critical point ($g \sim g_c$) can outperform those relying
solely on engineered quantum states of light. Thus, exploiting criticality-induced virtual squeezing provides a
promising route toward quantum-enhanced sensing that is both
practically viable and fundamentally distinct from conventional
approaches.}

\newpage
{\section{S4. Optimality of frequency measurements through the quadrature (homodyne) detection.}
In the main text, we have stated that in the case of single-mode Gaussian states, it is well
established that optimal measurements---those that saturate the quantum Cram\'er-Rao bound---are Gaussian, such as homodyne or
heterodyne detection (see Ref. [46] in the main text for details). Here, we show it explicitly by calculating the quantum Fisher information and the signal-to-noise ratio. For the sake of simplicity, we focus on the quantum Rabi model. In the basis of virtual excitations, the ground state
takes the simple form
\begin{align}
    |\psi_0\rangle = \hat S(\xi)|0\rangle,
\end{align}
where $\hat S(\xi)$ is the squeeze operator with the squeezing parameter $\xi = \frac{1}4\log\left(1-g^2/g_c^2\right)$ and $g_c = \sqrt{\omega \Omega}$. The quantum Fisher information becomes
\begin{align}
     \mathcal{I}_\omega = 4\left(\langle \partial_\omega \psi_0|\partial_\omega \psi_0\rangle  - |\langle \partial_\omega \psi_0|\psi_0\rangle|^2\right) = 2(\partial_\omega \xi)^2 = \frac{1}{8 \omega^2 \left(1 - \frac{g ^2}{g_c^2} \right)^2} \frac{g ^4}{g_c^4}
\end{align}
Now, let us calculate the signal-to-noise ratio for homodyne
detection, assuming that we are able to measure virtual
excitations. In this case, the signal-to-noise ratio can be evaluated using the
quadrature operators of the mode $\hat{a}$. Since the first moment
of the quadrature $\hat{X} = (\hat{a} + \hat{a}^\dagger)/2$
vanishes for a squeezed vacuum state, we must instead rely on the
second moment
\begin{align}
    \langle \hat X^2 \rangle = \frac{1}{4}\exp(- 2 \xi) = \frac{1}{4}\frac{1}{\sqrt{1-\frac{g^2}{\omega \Omega}}},
\end{align}
and the fourth moment,
\begin{align}
    \langle \hat X^4 \rangle = 3     \langle \hat X^2 \rangle^2 = \frac{3}{16}\exp(-4 \xi) = \frac{3}{16}\frac{1}{1 - \frac{g^2}{ \omega
    \Omega}}.
\end{align}
To evaluate the signal-to-noise ratio, we compute the derivative of the second
moment with respect to $\omega$. This yields the following
expression for the signal-to-noise ratio
\begin{align}
    S(\omega) = \frac{|\partial_\omega
    \langle \hat X^2\rangle|^2}{\langle \hat X^4\rangle - \langle \hat X^2 \rangle^2}
    = \frac{1}{8 \omega^2 \left(1 - \frac{g ^2}{g_c^2} \right)^2}
    \frac{g ^4}{g_c^4}. 
\end{align}
As expected, this result matches the quantum Fisher information
given in Eq.~(S27), even when considering measurements on virtual
excitations. This confirms that quadrature measurement is optimal
in this setting, as also discussed in detail in Ref.~[46] from the main text.

We now switch to the basis of measurable excitations. Since the
ground state contains no such excitations, we initialize the
system in a coherent state to introduce excitations. As a result,
the state of the system at time $t$ evolves to
\begin{align}
    |\psi(t)\rangle = |\alpha \exp(- i\omega_-t)\rangle \equiv |\alpha\exp(-i\omega e^{2\xi}t)\rangle.
\end{align}
The quantum Fisher information can then be readily calculated as
\begin{align}
    \mathcal I_\omega =  4|\alpha|^2 t^2 (\partial_\omega \omega_-)^2=|\alpha|^2 t^2 \frac{\left(g^2/g_c^2-2\right)^2}{ 1 -g^2/g_c^2}
\end{align}
We now turn to the signal-to-noise ratio for homodyne detection in this setting. Here, the signal-to-noise ratio is
calculated for the normal (measurable) mode $\hat{c}$ and its
associated quadrature $\hat{X}_c$. For the first moment, we
obtain:
\begin{align}
    \langle \hat X_c \rangle = |\alpha| \cos \left(\omega e^{2\xi} t \right),
\end{align}
and for the variance, we obtain
\begin{align}
    \langle \hat X_c^2 \rangle - \langle \hat X_c \rangle^2 = \frac{1}{4}.
\end{align}
By combining all the terms and taking the derivative with respect
to the unknown parameter $\omega$, we find that the signal-to-noise ratio is bounded
by
\begin{align}\label{eq:caseC}
    S(\omega) = |\alpha|^2 t^2  \frac{  \left(2-\frac{g^2}{g_c^2}\right)^2}{1-\frac{g^2}{g_c^2}} \sin^2  \left(\omega e^{2\xi} t\right).
\end{align}
By selecting the optimal quadrature angle---determined by the
state's orientation in phase space---we obtain
\begin{align}\label{eq:caseC}
    S(\omega) = |\alpha|^2 t^2  \frac{  \left({g^2}/{g_c^2}-2\right)^2}{1-{g^2}/{g_c^2}},
\end{align}
which is equal to the quantum Fisher information given in Eq.~(S32).
This proves that homodyne (quadrature) detection is optimal.


\begin{thebibliography}{74}%
\makeatletter
\providecommand \@ifxundefined [1]{%
 \@ifx{#1\undefined}
}%
\providecommand \@ifnum [1]{%
 \ifnum #1\expandafter \@firstoftwo
 \else \expandafter \@secondoftwo
 \fi
}%
\providecommand \@ifx [1]{%
 \ifx #1\expandafter \@firstoftwo
 \else \expandafter \@secondoftwo
 \fi
}%
\providecommand \natexlab [1]{#1}%
\providecommand \enquote  [1]{``#1''}%
\providecommand \bibnamefont  [1]{#1}%
\providecommand \bibfnamefont [1]{#1}%
\providecommand \citenamefont [1]{#1}%
\providecommand \href@noop [0]{\@secondoftwo}%
\providecommand \href [0]{\begingroup \@sanitize@url \@href}%
\providecommand \@href[1]{\@@startlink{#1}\@@href}%
\providecommand \@@href[1]{\endgroup#1\@@endlink}%
\providecommand \@sanitize@url [0]{\catcode `\\12\catcode `\$12\catcode `\&12\catcode `\#12\catcode `\^12\catcode `\_12\catcode `\%12\relax}%
\providecommand \@@startlink[1]{}%
\providecommand \@@endlink[0]{}%
\providecommand \url  [0]{\begingroup\@sanitize@url \@url }%
\providecommand \@url [1]{\endgroup\@href {#1}{\urlprefix }}%
\providecommand \urlprefix  [0]{URL }%
\providecommand \Eprint [0]{\href }%
\providecommand \doibase [0]{https://doi.org/}%
\providecommand \selectlanguage [0]{\@gobble}%
\providecommand \bibinfo  [0]{\@secondoftwo}%
\providecommand \bibfield  [0]{\@secondoftwo}%
\providecommand \translation [1]{[#1]}%
\providecommand \BibitemOpen [0]{}%
\providecommand \bibitemStop [0]{}%
\providecommand \bibitemNoStop [0]{.\EOS\space}%
\providecommand \EOS [0]{\spacefactor3000\relax}%
\providecommand \BibitemShut  [1]{\csname bibitem#1\endcsname}%
\let\auto@bib@innerbib\@empty
\bibitem [{\citenamefont {Frisk~Kockum}\ \emph {et~al.}(2019)\citenamefont {Frisk~Kockum}, \citenamefont {Miranowicz}, \citenamefont {De~Liberato}, \citenamefont {Savasta},\ and\ \citenamefont {Nori}}]{Nori2019USCreview}%
  \BibitemOpen
  \bibfield  {author} {\bibinfo {author} {\bibfnamefont {A.}~\bibnamefont {Frisk~Kockum}}, \bibinfo {author} {\bibfnamefont {A.}~\bibnamefont {Miranowicz}}, \bibinfo {author} {\bibfnamefont {S.}~\bibnamefont {De~Liberato}}, \bibinfo {author} {\bibfnamefont {S.}~\bibnamefont {Savasta}},\ and\ \bibinfo {author} {\bibfnamefont {F.}~\bibnamefont {Nori}},\ }\bibfield  {title} {\bibinfo {title} {Ultrastrong coupling between light and matter},\ }\href {https://doi.org/10.1038/s42254-018-0006-2} {\bibfield  {journal} {\bibinfo  {journal} {Nat Rev Phys}\ }\textbf {\bibinfo {volume} {1}},\ \bibinfo {pages} {19} (\bibinfo {year} {2019})}\BibitemShut {NoStop}%
\bibitem [{\citenamefont {Forn-D\'{\i}az}\ \emph {et~al.}(2019)\citenamefont {Forn-D\'{\i}az}, \citenamefont {Lamata}, \citenamefont {Rico}, \citenamefont {Kono},\ and\ \citenamefont {Solano}}]{solano2019rmp}%
  \BibitemOpen
  \bibfield  {author} {\bibinfo {author} {\bibfnamefont {P.}~\bibnamefont {Forn-D\'{\i}az}}, \bibinfo {author} {\bibfnamefont {L.}~\bibnamefont {Lamata}}, \bibinfo {author} {\bibfnamefont {E.}~\bibnamefont {Rico}}, \bibinfo {author} {\bibfnamefont {J.}~\bibnamefont {Kono}},\ and\ \bibinfo {author} {\bibfnamefont {E.}~\bibnamefont {Solano}},\ }\bibfield  {title} {\bibinfo {title} {Ultrastrong coupling regimes of light-matter interaction},\ }\href {https://doi.org/10.1103/RevModPhys.91.025005} {\bibfield  {journal} {\bibinfo  {journal} {Rev. Mod. Phys.}\ }\textbf {\bibinfo {volume} {91}},\ \bibinfo {pages} {025005} (\bibinfo {year} {2019})}\BibitemShut {NoStop}%
\bibitem [{\citenamefont {Qin}\ \emph {et~al.}(2024)\citenamefont {Qin}, \citenamefont {Kockum}, \citenamefont {Muñoz}, \citenamefont {Miranowicz},\ and\ \citenamefont {Nori}}]{QIN20241}%
  \BibitemOpen
  \bibfield  {author} {\bibinfo {author} {\bibfnamefont {W.}~\bibnamefont {Qin}}, \bibinfo {author} {\bibfnamefont {A.~F.}\ \bibnamefont {Kockum}}, \bibinfo {author} {\bibfnamefont {C.~S.}\ \bibnamefont {Muñoz}}, \bibinfo {author} {\bibfnamefont {A.}~\bibnamefont {Miranowicz}},\ and\ \bibinfo {author} {\bibfnamefont {F.}~\bibnamefont {Nori}},\ }\bibfield  {title} {\bibinfo {title} {Quantum amplification and simulation of strong and ultrastrong coupling of light and matter},\ }\href {https://doi.org/https://doi.org/10.1016/j.physrep.2024.05.003} {\bibfield  {journal} {\bibinfo  {journal} {Phys. Rep.}\ }\textbf {\bibinfo {volume} {1078}},\ \bibinfo {pages} {1} (\bibinfo {year} {2024})}\BibitemShut {NoStop}%
\bibitem [{\citenamefont {MacFarlane}\ \emph {et~al.}(2003)\citenamefont {MacFarlane}, \citenamefont {Dowling},\ and\ \citenamefont {Milburn}}]{quantumtechnologies2003Milburn}%
  \BibitemOpen
  \bibfield  {author} {\bibinfo {author} {\bibfnamefont {A.~G.~J.}\ \bibnamefont {MacFarlane}}, \bibinfo {author} {\bibfnamefont {J.~P.}\ \bibnamefont {Dowling}},\ and\ \bibinfo {author} {\bibfnamefont {G.~J.}\ \bibnamefont {Milburn}},\ }\bibfield  {title} {\bibinfo {title} {Quantum technology: the second quantum revolution},\ }\href {https://doi.org/10.1098/rsta.2003.1227} {\bibfield  {journal} {\bibinfo  {journal} {Philos. Trans. R. Soc. A}\ }\textbf {\bibinfo {volume} {361}},\ \bibinfo {pages} {1655} (\bibinfo {year} {2003})}\BibitemShut {NoStop}%
\bibitem [{\citenamefont {Kurizki}\ \emph {et~al.}(2015)\citenamefont {Kurizki}, \citenamefont {Bertet}, \citenamefont {Kubo}, \citenamefont {Mølmer}, \citenamefont {Petrosyan}, \citenamefont {Rabl},\ and\ \citenamefont {Schmiedmayer}}]{QT2015Schmiedmayer}%
  \BibitemOpen
  \bibfield  {author} {\bibinfo {author} {\bibfnamefont {G.}~\bibnamefont {Kurizki}}, \bibinfo {author} {\bibfnamefont {P.}~\bibnamefont {Bertet}}, \bibinfo {author} {\bibfnamefont {Y.}~\bibnamefont {Kubo}}, \bibinfo {author} {\bibfnamefont {K.}~\bibnamefont {Mølmer}}, \bibinfo {author} {\bibfnamefont {D.}~\bibnamefont {Petrosyan}}, \bibinfo {author} {\bibfnamefont {P.}~\bibnamefont {Rabl}},\ and\ \bibinfo {author} {\bibfnamefont {J.}~\bibnamefont {Schmiedmayer}},\ }\bibfield  {title} {\bibinfo {title} {Quantum technologies with hybrid systems},\ }\href {https://doi.org/10.1073/pnas.1419326112} {\bibfield  {journal} {\bibinfo  {journal} {Proc. Natl. Acad. Sci. U.S.A.}\ }\textbf {\bibinfo {volume} {112}},\ \bibinfo {pages} {3866} (\bibinfo {year} {2015})}\BibitemShut {NoStop}%
\bibitem [{\citenamefont {Ciuti}\ \emph {et~al.}(2005)\citenamefont {Ciuti}, \citenamefont {Bastard},\ and\ \citenamefont {Carusotto}}]{bastard2005vacuumproperties}%
  \BibitemOpen
  \bibfield  {author} {\bibinfo {author} {\bibfnamefont {C.}~\bibnamefont {Ciuti}}, \bibinfo {author} {\bibfnamefont {G.}~\bibnamefont {Bastard}},\ and\ \bibinfo {author} {\bibfnamefont {I.}~\bibnamefont {Carusotto}},\ }\bibfield  {title} {\bibinfo {title} {Quantum vacuum properties of the intersubband cavity polariton field},\ }\href {https://doi.org/10.1103/PhysRevB.72.115303} {\bibfield  {journal} {\bibinfo  {journal} {Phys. Rev. B}\ }\textbf {\bibinfo {volume} {72}},\ \bibinfo {pages} {115303} (\bibinfo {year} {2005})}\BibitemShut {NoStop}%
\bibitem [{\citenamefont {Ciuti}\ and\ \citenamefont {Carusotto}(2006)}]{Ciuti2006inputoutputUSC}%
  \BibitemOpen
  \bibfield  {author} {\bibinfo {author} {\bibfnamefont {C.}~\bibnamefont {Ciuti}}\ and\ \bibinfo {author} {\bibfnamefont {I.}~\bibnamefont {Carusotto}},\ }\bibfield  {title} {\bibinfo {title} {Input-output theory of cavities in the ultrastrong coupling regime: {T}he case of time-independent cavity parameters},\ }\href {https://doi.org/10.1103/PhysRevA.74.033811} {\bibfield  {journal} {\bibinfo  {journal} {Phys. Rev. A}\ }\textbf {\bibinfo {volume} {74}},\ \bibinfo {pages} {033811} (\bibinfo {year} {2006})}\BibitemShut {NoStop}%
\bibitem [{\citenamefont {Anappara}\ \emph {et~al.}(2009)\citenamefont {Anappara}, \citenamefont {De~Liberato}, \citenamefont {Tredicucci}, \citenamefont {Ciuti}, \citenamefont {Biasiol}, \citenamefont {Sorba},\ and\ \citenamefont {Beltram}}]{Ciuti2009signaturesUSC}%
  \BibitemOpen
  \bibfield  {author} {\bibinfo {author} {\bibfnamefont {A.~A.}\ \bibnamefont {Anappara}}, \bibinfo {author} {\bibfnamefont {S.}~\bibnamefont {De~Liberato}}, \bibinfo {author} {\bibfnamefont {A.}~\bibnamefont {Tredicucci}}, \bibinfo {author} {\bibfnamefont {C.}~\bibnamefont {Ciuti}}, \bibinfo {author} {\bibfnamefont {G.}~\bibnamefont {Biasiol}}, \bibinfo {author} {\bibfnamefont {L.}~\bibnamefont {Sorba}},\ and\ \bibinfo {author} {\bibfnamefont {F.}~\bibnamefont {Beltram}},\ }\bibfield  {title} {\bibinfo {title} {Signatures of the ultrastrong light-matter coupling regime},\ }\href {https://doi.org/10.1103/PhysRevB.79.201303} {\bibfield  {journal} {\bibinfo  {journal} {Phys. Rev. B}\ }\textbf {\bibinfo {volume} {79}},\ \bibinfo {pages} {201303} (\bibinfo {year} {2009})}\BibitemShut {NoStop}%
\bibitem [{\citenamefont {Todorov}\ \emph {et~al.}(2010)\citenamefont {Todorov}, \citenamefont {Andrews}, \citenamefont {Colombelli}, \citenamefont {De~Liberato}, \citenamefont {Ciuti}, \citenamefont {Klang}, \citenamefont {Strasser},\ and\ \citenamefont {Sirtori}}]{Todorov2010USCpolaritondots}%
  \BibitemOpen
  \bibfield  {author} {\bibinfo {author} {\bibfnamefont {Y.}~\bibnamefont {Todorov}}, \bibinfo {author} {\bibfnamefont {A.~M.}\ \bibnamefont {Andrews}}, \bibinfo {author} {\bibfnamefont {R.}~\bibnamefont {Colombelli}}, \bibinfo {author} {\bibfnamefont {S.}~\bibnamefont {De~Liberato}}, \bibinfo {author} {\bibfnamefont {C.}~\bibnamefont {Ciuti}}, \bibinfo {author} {\bibfnamefont {P.}~\bibnamefont {Klang}}, \bibinfo {author} {\bibfnamefont {G.}~\bibnamefont {Strasser}},\ and\ \bibinfo {author} {\bibfnamefont {C.}~\bibnamefont {Sirtori}},\ }\bibfield  {title} {\bibinfo {title} {Ultrastrong {L}ight-{M}atter {C}oupling {R}egime with {P}olariton {D}ots},\ }\href {https://doi.org/10.1103/PhysRevLett.105.196402} {\bibfield  {journal} {\bibinfo  {journal} {Phys. Rev. Lett.}\ }\textbf {\bibinfo {volume} {105}},\ \bibinfo {pages} {196402} (\bibinfo {year} {2010})}\BibitemShut {NoStop}%
\bibitem [{\citenamefont {Ashhab}\ and\ \citenamefont {Nori}(2010)}]{Nori2010qubitoscillatorUSC}%
  \BibitemOpen
  \bibfield  {author} {\bibinfo {author} {\bibfnamefont {S.}~\bibnamefont {Ashhab}}\ and\ \bibinfo {author} {\bibfnamefont {F.}~\bibnamefont {Nori}},\ }\bibfield  {title} {\bibinfo {title} {Qubit-oscillator systems in the ultrastrong-coupling regime and their potential for preparing nonclassical states},\ }\href {https://doi.org/10.1103/PhysRevA.81.042311} {\bibfield  {journal} {\bibinfo  {journal} {Phys. Rev. A}\ }\textbf {\bibinfo {volume} {81}},\ \bibinfo {pages} {042311} (\bibinfo {year} {2010})}\BibitemShut {NoStop}%
\bibitem [{\citenamefont {Beaudoin}\ \emph {et~al.}(2011)\citenamefont {Beaudoin}, \citenamefont {Gambetta},\ and\ \citenamefont {Blais}}]{blais2011dissipationUSC}%
  \BibitemOpen
  \bibfield  {author} {\bibinfo {author} {\bibfnamefont {F.}~\bibnamefont {Beaudoin}}, \bibinfo {author} {\bibfnamefont {J.~M.}\ \bibnamefont {Gambetta}},\ and\ \bibinfo {author} {\bibfnamefont {A.}~\bibnamefont {Blais}},\ }\bibfield  {title} {\bibinfo {title} {Dissipation and ultrastrong coupling in circuit {Q}{E}{D}},\ }\href {https://doi.org/10.1103/PhysRevA.84.043832} {\bibfield  {journal} {\bibinfo  {journal} {Phys. Rev. A}\ }\textbf {\bibinfo {volume} {84}},\ \bibinfo {pages} {043832} (\bibinfo {year} {2011})}\BibitemShut {NoStop}%
\bibitem [{\citenamefont {Settineri}\ \emph {et~al.}(2018)\citenamefont {Settineri}, \citenamefont {Macr\'{\i}}, \citenamefont {Ridolfo}, \citenamefont {Di~Stefano}, \citenamefont {Kockum}, \citenamefont {Nori},\ and\ \citenamefont {Savasta}}]{savasta2018dissipationUSC}%
  \BibitemOpen
  \bibfield  {author} {\bibinfo {author} {\bibfnamefont {A.}~\bibnamefont {Settineri}}, \bibinfo {author} {\bibfnamefont {V.}~\bibnamefont {Macr\'{\i}}}, \bibinfo {author} {\bibfnamefont {A.}~\bibnamefont {Ridolfo}}, \bibinfo {author} {\bibfnamefont {O.}~\bibnamefont {Di~Stefano}}, \bibinfo {author} {\bibfnamefont {A.~F.}\ \bibnamefont {Kockum}}, \bibinfo {author} {\bibfnamefont {F.}~\bibnamefont {Nori}},\ and\ \bibinfo {author} {\bibfnamefont {S.}~\bibnamefont {Savasta}},\ }\bibfield  {title} {\bibinfo {title} {Dissipation and thermal noise in hybrid quantum systems in the ultrastrong-coupling regime},\ }\href {https://doi.org/10.1103/PhysRevA.98.053834} {\bibfield  {journal} {\bibinfo  {journal} {Phys. Rev. A}\ }\textbf {\bibinfo {volume} {98}},\ \bibinfo {pages} {053834} (\bibinfo {year} {2018})}\BibitemShut {NoStop}%
\bibitem [{\citenamefont {Peterson}\ \emph {et~al.}(2019)\citenamefont {Peterson}, \citenamefont {Kotler}, \citenamefont {Lecocq}, \citenamefont {Cicak}, \citenamefont {Jin}, \citenamefont {Simmonds}, \citenamefont {Aumentado},\ and\ \citenamefont {Teufel}}]{Teufel2019Ulstrastrongmechanicalcavity}%
  \BibitemOpen
  \bibfield  {author} {\bibinfo {author} {\bibfnamefont {G.~A.}\ \bibnamefont {Peterson}}, \bibinfo {author} {\bibfnamefont {S.}~\bibnamefont {Kotler}}, \bibinfo {author} {\bibfnamefont {F.}~\bibnamefont {Lecocq}}, \bibinfo {author} {\bibfnamefont {K.}~\bibnamefont {Cicak}}, \bibinfo {author} {\bibfnamefont {X.~Y.}\ \bibnamefont {Jin}}, \bibinfo {author} {\bibfnamefont {R.~W.}\ \bibnamefont {Simmonds}}, \bibinfo {author} {\bibfnamefont {J.}~\bibnamefont {Aumentado}},\ and\ \bibinfo {author} {\bibfnamefont {J.~D.}\ \bibnamefont {Teufel}},\ }\bibfield  {title} {\bibinfo {title} {Ultrastrong {P}arametric {C}oupling between a {S}uperconducting {C}avity and a {M}echanical {R}esonator},\ }\href {https://doi.org/10.1103/PhysRevLett.123.247701} {\bibfield  {journal} {\bibinfo  {journal} {Phys. Rev. Lett.}\ }\textbf {\bibinfo {volume} {123}},\ \bibinfo {pages} {247701} (\bibinfo {year} {2019})}\BibitemShut {NoStop}%
\bibitem [{\citenamefont {Bibak}\ \emph {et~al.}(2023)\citenamefont {Bibak}, \citenamefont {Deli\ifmmode~\acute{c}\else \'{c}\fi{}}, \citenamefont {Aspelmeyer},\ and\ \citenamefont {Daki\ifmmode~\acute{c}\else \'{c}\fi{}}}]{uros2023dissipativephasetrans}%
  \BibitemOpen
  \bibfield  {author} {\bibinfo {author} {\bibfnamefont {F.}~\bibnamefont {Bibak}}, \bibinfo {author} {\bibfnamefont {U.}~\bibnamefont {Deli\ifmmode~\acute{c}\else \'{c}\fi{}}}, \bibinfo {author} {\bibfnamefont {M.}~\bibnamefont {Aspelmeyer}},\ and\ \bibinfo {author} {\bibfnamefont {B.}~\bibnamefont {Daki\ifmmode~\acute{c}\else \'{c}\fi{}}},\ }\bibfield  {title} {\bibinfo {title} {Dissipative phase transitions in optomechanical systems},\ }\href {https://doi.org/10.1103/PhysRevA.107.053505} {\bibfield  {journal} {\bibinfo  {journal} {Phys. Rev. A}\ }\textbf {\bibinfo {volume} {107}},\ \bibinfo {pages} {053505} (\bibinfo {year} {2023})}\BibitemShut {NoStop}%
\bibitem [{\citenamefont {Gietka}\ \emph {et~al.}(2023)\citenamefont {Gietka}, \citenamefont {Hotter},\ and\ \citenamefont {Ritsch}}]{gietks2023USquezingQRM}%
  \BibitemOpen
  \bibfield  {author} {\bibinfo {author} {\bibfnamefont {K.}~\bibnamefont {Gietka}}, \bibinfo {author} {\bibfnamefont {C.}~\bibnamefont {Hotter}},\ and\ \bibinfo {author} {\bibfnamefont {H.}~\bibnamefont {Ritsch}},\ }\bibfield  {title} {\bibinfo {title} {Unique {S}teady-{S}tate {S}queezing in a {D}riven {Q}uantum {R}abi {M}odel},\ }\href {https://doi.org/10.1103/PhysRevLett.131.223604} {\bibfield  {journal} {\bibinfo  {journal} {Phys. Rev. Lett.}\ }\textbf {\bibinfo {volume} {131}},\ \bibinfo {pages} {223604} (\bibinfo {year} {2023})}\BibitemShut {NoStop}%
\bibitem [{\citenamefont {Stassi}\ \emph {et~al.}(2023)\citenamefont {Stassi}, \citenamefont {Cirio}, \citenamefont {Funo}, \citenamefont {Puebla}, \citenamefont {Lambert},\ and\ \citenamefont {Nori}}]{stassi2023unvelingveiling}%
  \BibitemOpen
  \bibfield  {author} {\bibinfo {author} {\bibfnamefont {R.}~\bibnamefont {Stassi}}, \bibinfo {author} {\bibfnamefont {M.}~\bibnamefont {Cirio}}, \bibinfo {author} {\bibfnamefont {K.}~\bibnamefont {Funo}}, \bibinfo {author} {\bibfnamefont {J.}~\bibnamefont {Puebla}}, \bibinfo {author} {\bibfnamefont {N.}~\bibnamefont {Lambert}},\ and\ \bibinfo {author} {\bibfnamefont {F.}~\bibnamefont {Nori}},\ }\bibfield  {title} {\bibinfo {title} {Unveiling and veiling an entangled light-matter quantum state from the vacuum},\ }\href {https://doi.org/10.1103/PhysRevResearch.5.043095} {\bibfield  {journal} {\bibinfo  {journal} {Phys. Rev. Res.}\ }\textbf {\bibinfo {volume} {5}},\ \bibinfo {pages} {043095} (\bibinfo {year} {2023})}\BibitemShut {NoStop}%
\bibitem [{\citenamefont {Dare}\ \emph {et~al.}(2024{\natexlab{a}})\citenamefont {Dare}, \citenamefont {Hansen}, \citenamefont {Coroli}, \citenamefont {Johnson}, \citenamefont {Aspelmeyer},\ and\ \citenamefont {Deli\ifmmode~\acute{c}\else \'{c}\fi{}}}]{UrosDelic2024OPtomochenicalUSC}%
  \BibitemOpen
  \bibfield  {author} {\bibinfo {author} {\bibfnamefont {K.}~\bibnamefont {Dare}}, \bibinfo {author} {\bibfnamefont {J.~J.}\ \bibnamefont {Hansen}}, \bibinfo {author} {\bibfnamefont {I.}~\bibnamefont {Coroli}}, \bibinfo {author} {\bibfnamefont {A.}~\bibnamefont {Johnson}}, \bibinfo {author} {\bibfnamefont {M.}~\bibnamefont {Aspelmeyer}},\ and\ \bibinfo {author} {\bibfnamefont {U.}~\bibnamefont {Deli\ifmmode~\acute{c}\else \'{c}\fi{}}},\ }\bibfield  {title} {\bibinfo {title} {Ultrastrong linear optomechanical interaction},\ }\href {https://doi.org/10.1103/PhysRevResearch.6.L042025} {\bibfield  {journal} {\bibinfo  {journal} {Phys. Rev. Res.}\ }\textbf {\bibinfo {volume} {6}},\ \bibinfo {pages} {L042025} (\bibinfo {year} {2024}{\natexlab{a}})}\BibitemShut {NoStop}%
\bibitem [{\citenamefont {Chen}\ \emph {et~al.}(2024)\citenamefont {Chen}, \citenamefont {Shi}, \citenamefont {Zhang}, \citenamefont {Nori},\ and\ \citenamefont {Xia}}]{chen2024suppressedenergyrelaxationquantum}%
  \BibitemOpen
  \bibfield  {author} {\bibinfo {author} {\bibfnamefont {Y.-H.}\ \bibnamefont {Chen}}, \bibinfo {author} {\bibfnamefont {Z.-C.}\ \bibnamefont {Shi}}, \bibinfo {author} {\bibfnamefont {Y.-R.}\ \bibnamefont {Zhang}}, \bibinfo {author} {\bibfnamefont {F.}~\bibnamefont {Nori}},\ and\ \bibinfo {author} {\bibfnamefont {Y.}~\bibnamefont {Xia}},\ }\href {https://arxiv.org/abs/2411.03710} {\bibinfo {title} {Suppressed {E}nergy {R}elaxation in the {Q}uantum {R}abi {M}odel at the {C}ritical {P}oint}} (\bibinfo {year} {2024}),\ \Eprint {https://arxiv.org/abs/2411.03710} {arXiv:2411.03710 [quant-ph]} \BibitemShut {NoStop}%
\bibitem [{\citenamefont {Cao}\ \emph {et~al.}(2011)\citenamefont {Cao}, \citenamefont {You}, \citenamefont {Zheng},\ and\ \citenamefont {Nori}}]{Cao_2011Noriassymetry}%
  \BibitemOpen
  \bibfield  {author} {\bibinfo {author} {\bibfnamefont {X.}~\bibnamefont {Cao}}, \bibinfo {author} {\bibfnamefont {J.~Q.}\ \bibnamefont {You}}, \bibinfo {author} {\bibfnamefont {H.}~\bibnamefont {Zheng}},\ and\ \bibinfo {author} {\bibfnamefont {F.}~\bibnamefont {Nori}},\ }\bibfield  {title} {\bibinfo {title} {A qubit strongly coupled to a resonant cavity: asymmetry of the spontaneous emission spectrum beyond the rotating wave approximation},\ }\href {https://doi.org/10.1088/1367-2630/13/7/073002} {\bibfield  {journal} {\bibinfo  {journal} {New J. Phys.}\ }\textbf {\bibinfo {volume} {13}},\ \bibinfo {pages} {073002} (\bibinfo {year} {2011})}\BibitemShut {NoStop}%
\bibitem [{\citenamefont {Gietka}(2024)}]{gietka2024vacuumrabisplittingmanifestation}%
  \BibitemOpen
  \bibfield  {author} {\bibinfo {author} {\bibfnamefont {K.}~\bibnamefont {Gietka}},\ }\bibfield  {title} {\bibinfo {title} {Vacuum {R}abi splitting as a manifestation of virtual two-mode squeezing: {E}xtracting the squeezing parameters from frequency shifts},\ }\href {https://doi.org/10.1103/PhysRevA.110.063703} {\bibfield  {journal} {\bibinfo  {journal} {Phys. Rev. A}\ }\textbf {\bibinfo {volume} {110}},\ \bibinfo {pages} {063703} (\bibinfo {year} {2024})}\BibitemShut {NoStop}%
\bibitem [{\citenamefont {Garraway}(2011)}]{garraway2011dickerevisisted}%
  \BibitemOpen
  \bibfield  {author} {\bibinfo {author} {\bibfnamefont {B.~M.}\ \bibnamefont {Garraway}},\ }\bibfield  {title} {\bibinfo {title} {The dicke model in quantum optics: Dicke model revisited},\ }\href {https://doi.org/10.1098/rsta.2010.0333} {\bibfield  {journal} {\bibinfo  {journal} {Philos. Trans. R. Soc. A}\ }\textbf {\bibinfo {volume} {369}},\ \bibinfo {pages} {1137} (\bibinfo {year} {2011})}\BibitemShut {NoStop}%
\bibitem [{\citenamefont {Ritsch}\ \emph {et~al.}(2013)\citenamefont {Ritsch}, \citenamefont {Domokos}, \citenamefont {Brennecke},\ and\ \citenamefont {Esslinger}}]{helmut2013rmp}%
  \BibitemOpen
  \bibfield  {author} {\bibinfo {author} {\bibfnamefont {H.}~\bibnamefont {Ritsch}}, \bibinfo {author} {\bibfnamefont {P.}~\bibnamefont {Domokos}}, \bibinfo {author} {\bibfnamefont {F.}~\bibnamefont {Brennecke}},\ and\ \bibinfo {author} {\bibfnamefont {T.}~\bibnamefont {Esslinger}},\ }\bibfield  {title} {\bibinfo {title} {Cold atoms in cavity-generated dynamical optical potentials},\ }\href {https://doi.org/10.1103/RevModPhys.85.553} {\bibfield  {journal} {\bibinfo  {journal} {Rev. Mod. Phys.}\ }\textbf {\bibinfo {volume} {85}},\ \bibinfo {pages} {553} (\bibinfo {year} {2013})}\BibitemShut {NoStop}%
\bibitem [{\citenamefont {Emary}\ and\ \citenamefont {Brandes}(2003)}]{emarybrandes2003dickechaosphase}%
  \BibitemOpen
  \bibfield  {author} {\bibinfo {author} {\bibfnamefont {C.}~\bibnamefont {Emary}}\ and\ \bibinfo {author} {\bibfnamefont {T.}~\bibnamefont {Brandes}},\ }\bibfield  {title} {\bibinfo {title} {Chaos and the quantum phase transition in the {D}icke model},\ }\href {https://doi.org/10.1103/PhysRevE.67.066203} {\bibfield  {journal} {\bibinfo  {journal} {Phys. Rev. E}\ }\textbf {\bibinfo {volume} {67}},\ \bibinfo {pages} {066203} (\bibinfo {year} {2003})}\BibitemShut {NoStop}%
\bibitem [{\citenamefont {Larson}\ and\ \citenamefont {Irish}(2017)}]{Larson_2017}%
  \BibitemOpen
  \bibfield  {author} {\bibinfo {author} {\bibfnamefont {J.}~\bibnamefont {Larson}}\ and\ \bibinfo {author} {\bibfnamefont {E.~K.}\ \bibnamefont {Irish}},\ }\bibfield  {title} {\bibinfo {title} {Some remarks on ‘superradiant’ phase transitions in light-matter systems},\ }\href {https://doi.org/10.1088/1751-8121/aa65dc} {\bibfield  {journal} {\bibinfo  {journal} {J. Phys. A-Math.}\ }\textbf {\bibinfo {volume} {50}},\ \bibinfo {pages} {174002} (\bibinfo {year} {2017})}\BibitemShut {NoStop}%
\bibitem [{\citenamefont {Zhou}\ \emph {et~al.}(2020)\citenamefont {Zhou}, \citenamefont {Zhou}, \citenamefont {Yin}, \citenamefont {Huang},\ and\ \citenamefont {Liao}}]{zhou2HO2020}%
  \BibitemOpen
  \bibfield  {author} {\bibinfo {author} {\bibfnamefont {J.-Y.}\ \bibnamefont {Zhou}}, \bibinfo {author} {\bibfnamefont {Y.-H.}\ \bibnamefont {Zhou}}, \bibinfo {author} {\bibfnamefont {X.-L.}\ \bibnamefont {Yin}}, \bibinfo {author} {\bibfnamefont {J.-F.}\ \bibnamefont {Huang}},\ and\ \bibinfo {author} {\bibfnamefont {J.-Q.}\ \bibnamefont {Liao}},\ }\bibfield  {title} {\bibinfo {title} {Quantum entanglement maintained by virtual excitations in an ultrastrongly-coupled-oscillator system},\ }\href {https://doi.org/10.1038/s41598-020-68309-3} {\bibfield  {journal} {\bibinfo  {journal} {Scientific Reports}\ }\textbf {\bibinfo {volume} {10}},\ \bibinfo {pages} {12557} (\bibinfo {year} {2020})}\BibitemShut {NoStop}%
\bibitem [{\citenamefont {Basov}\ \emph {et~al.}(2021)\citenamefont {Basov}, \citenamefont {Asenjo-Garcia}, \citenamefont {Schuck}, \citenamefont {Zhu},\ and\ \citenamefont {Rubio}}]{BasovAsenjoGarciaSchuckZhuRubio2021}%
  \BibitemOpen
  \bibfield  {author} {\bibinfo {author} {\bibfnamefont {D.~N.}\ \bibnamefont {Basov}}, \bibinfo {author} {\bibfnamefont {A.}~\bibnamefont {Asenjo-Garcia}}, \bibinfo {author} {\bibfnamefont {P.~J.}\ \bibnamefont {Schuck}}, \bibinfo {author} {\bibfnamefont {X.}~\bibnamefont {Zhu}},\ and\ \bibinfo {author} {\bibfnamefont {A.}~\bibnamefont {Rubio}},\ }\bibfield  {title} {\bibinfo {title} {Polariton panorama},\ }\href {https://doi.org/doi:10.1515/nanoph-2020-0449} {\bibfield  {journal} {\bibinfo  {journal} {Nanophotonics}\ }\textbf {\bibinfo {volume} {10}},\ \bibinfo {pages} {549} (\bibinfo {year} {2021})}\BibitemShut {NoStop}%
\bibitem [{\citenamefont {Wilson}\ \emph {et~al.}(2011)\citenamefont {Wilson}, \citenamefont {Johansson}, \citenamefont {Pourkabirian}, \citenamefont {Simoen}, \citenamefont {Johansson}, \citenamefont {Duty}, \citenamefont {Nori},\ and\ \citenamefont {Delsing}}]{nori2011dynamiclcaeff}%
  \BibitemOpen
  \bibfield  {author} {\bibinfo {author} {\bibfnamefont {C.~M.}\ \bibnamefont {Wilson}}, \bibinfo {author} {\bibfnamefont {G.}~\bibnamefont {Johansson}}, \bibinfo {author} {\bibfnamefont {A.}~\bibnamefont {Pourkabirian}}, \bibinfo {author} {\bibfnamefont {M.}~\bibnamefont {Simoen}}, \bibinfo {author} {\bibfnamefont {J.~R.}\ \bibnamefont {Johansson}}, \bibinfo {author} {\bibfnamefont {T.}~\bibnamefont {Duty}}, \bibinfo {author} {\bibfnamefont {F.}~\bibnamefont {Nori}},\ and\ \bibinfo {author} {\bibfnamefont {P.}~\bibnamefont {Delsing}},\ }\bibfield  {title} {\bibinfo {title} {Observation of the dynamical {C}asimir effect in a superconducting circuit},\ }\href {https://doi.org/10.1038/nature10561} {\bibfield  {journal} {\bibinfo  {journal} {Nature}\ }\textbf {\bibinfo {volume} {479}},\ \bibinfo {pages} {376} (\bibinfo {year} {2011})}\BibitemShut {NoStop}%
\bibitem [{\citenamefont {Dodonov}(2020)}]{physics2010007}%
  \BibitemOpen
  \bibfield  {author} {\bibinfo {author} {\bibfnamefont {V.}~\bibnamefont {Dodonov}},\ }\bibfield  {title} {\bibinfo {title} {Fifty {Y}ears of the {D}ynamical {C}asimir {E}ffect},\ }\href {https://doi.org/10.3390/physics2010007} {\bibfield  {journal} {\bibinfo  {journal} {Physics}\ }\textbf {\bibinfo {volume} {2}},\ \bibinfo {pages} {67} (\bibinfo {year} {2020})}\BibitemShut {NoStop}%
\bibitem [{sup()}]{sup1}%
  \BibitemOpen
  \href@noop {} {}\bibinfo {howpublished} {\url{URL_will_be_inserted_by_publisher}}\BibitemShut {NoStop}%
\bibitem [{\citenamefont {De~Liberato}(2017)}]{DeLiberato2017virtualphotonosdisspative}%
  \BibitemOpen
  \bibfield  {author} {\bibinfo {author} {\bibfnamefont {S.}~\bibnamefont {De~Liberato}},\ }\bibfield  {title} {\bibinfo {title} {Virtual photons in the ground state of a dissipative system},\ }\href {https://doi.org/10.1038/s41467-017-01504-5} {\bibfield  {journal} {\bibinfo  {journal} {Nat. Commun.}\ }\textbf {\bibinfo {volume} {8}},\ \bibinfo {pages} {1465} (\bibinfo {year} {2017})}\BibitemShut {NoStop}%
\bibitem [{\citenamefont {Campos~Venuti}\ and\ \citenamefont {Zanardi}(2007)}]{zanardi2007criticaltensors}%
  \BibitemOpen
  \bibfield  {author} {\bibinfo {author} {\bibfnamefont {L.}~\bibnamefont {Campos~Venuti}}\ and\ \bibinfo {author} {\bibfnamefont {P.}~\bibnamefont {Zanardi}},\ }\bibfield  {title} {\bibinfo {title} {Quantum {C}ritical {S}caling of the {G}eometric {T}ensors},\ }\href {https://doi.org/10.1103/PhysRevLett.99.095701} {\bibfield  {journal} {\bibinfo  {journal} {Phys. Rev. Lett.}\ }\textbf {\bibinfo {volume} {99}},\ \bibinfo {pages} {095701} (\bibinfo {year} {2007})}\BibitemShut {NoStop}%
\bibitem [{\citenamefont {Salvatori}\ \emph {et~al.}(2014)\citenamefont {Salvatori}, \citenamefont {Mandarino},\ and\ \citenamefont {Paris}}]{paris2014LMGcrit}%
  \BibitemOpen
  \bibfield  {author} {\bibinfo {author} {\bibfnamefont {G.}~\bibnamefont {Salvatori}}, \bibinfo {author} {\bibfnamefont {A.}~\bibnamefont {Mandarino}},\ and\ \bibinfo {author} {\bibfnamefont {M.~G.~A.}\ \bibnamefont {Paris}},\ }\bibfield  {title} {\bibinfo {title} {Quantum metrology in {L}ipkin-{M}eshkov-{G}lick critical systems},\ }\href {https://doi.org/10.1103/PhysRevA.90.022111} {\bibfield  {journal} {\bibinfo  {journal} {Phys. Rev. A}\ }\textbf {\bibinfo {volume} {90}},\ \bibinfo {pages} {022111} (\bibinfo {year} {2014})}\BibitemShut {NoStop}%
\bibitem [{\citenamefont {Rams}\ \emph {et~al.}(2018)\citenamefont {Rams}, \citenamefont {Sierant}, \citenamefont {Dutta}, \citenamefont {Horodecki},\ and\ \citenamefont {Zakrzewski}}]{zakrzewski2018QCMlimits}%
  \BibitemOpen
  \bibfield  {author} {\bibinfo {author} {\bibfnamefont {M.~M.}\ \bibnamefont {Rams}}, \bibinfo {author} {\bibfnamefont {P.}~\bibnamefont {Sierant}}, \bibinfo {author} {\bibfnamefont {O.}~\bibnamefont {Dutta}}, \bibinfo {author} {\bibfnamefont {P.}~\bibnamefont {Horodecki}},\ and\ \bibinfo {author} {\bibfnamefont {J.}~\bibnamefont {Zakrzewski}},\ }\bibfield  {title} {\bibinfo {title} {At the {L}imits of {C}riticality-{B}ased {Q}uantum {M}etrology: {A}pparent {S}uper-{H}eisenberg {S}caling {R}evisited},\ }\href {https://doi.org/10.1103/PhysRevX.8.021022} {\bibfield  {journal} {\bibinfo  {journal} {Phys. Rev. X}\ }\textbf {\bibinfo {volume} {8}},\ \bibinfo {pages} {021022} (\bibinfo {year} {2018})}\BibitemShut {NoStop}%
\bibitem [{\citenamefont {Garbe}\ \emph {et~al.}(2020)\citenamefont {Garbe}, \citenamefont {Bina}, \citenamefont {Keller}, \citenamefont {Paris},\ and\ \citenamefont {Felicetti}}]{garbe2020criticalmet}%
  \BibitemOpen
  \bibfield  {author} {\bibinfo {author} {\bibfnamefont {L.}~\bibnamefont {Garbe}}, \bibinfo {author} {\bibfnamefont {M.}~\bibnamefont {Bina}}, \bibinfo {author} {\bibfnamefont {A.}~\bibnamefont {Keller}}, \bibinfo {author} {\bibfnamefont {M.~G.~A.}\ \bibnamefont {Paris}},\ and\ \bibinfo {author} {\bibfnamefont {S.}~\bibnamefont {Felicetti}},\ }\bibfield  {title} {\bibinfo {title} {Critical {Q}uantum {M}etrology with a {F}inite-{C}omponent {Q}uantum {P}hase {T}ransition},\ }\href {https://doi.org/10.1103/PhysRevLett.124.120504} {\bibfield  {journal} {\bibinfo  {journal} {Phys. Rev. Lett.}\ }\textbf {\bibinfo {volume} {124}},\ \bibinfo {pages} {120504} (\bibinfo {year} {2020})}\BibitemShut {NoStop}%
\bibitem [{\citenamefont {Gietka}\ \emph {et~al.}(2021)\citenamefont {Gietka}, \citenamefont {Metz}, \citenamefont {Keller},\ and\ \citenamefont {Li}}]{Gietka2021adiabaticcritical}%
  \BibitemOpen
  \bibfield  {author} {\bibinfo {author} {\bibfnamefont {K.}~\bibnamefont {Gietka}}, \bibinfo {author} {\bibfnamefont {F.}~\bibnamefont {Metz}}, \bibinfo {author} {\bibfnamefont {T.}~\bibnamefont {Keller}},\ and\ \bibinfo {author} {\bibfnamefont {J.}~\bibnamefont {Li}},\ }\bibfield  {title} {\bibinfo {title} {Adiabatic critical quantum metrology cannot reach the {H}eisenberg limit even when shortcuts to adiabaticity are applied},\ }\href {https://doi.org/10.22331/q-2021-07-01-489} {\bibfield  {journal} {\bibinfo  {journal} {{Quantum}}\ }\textbf {\bibinfo {volume} {5}},\ \bibinfo {pages} {489} (\bibinfo {year} {2021})}\BibitemShut {NoStop}%
\bibitem [{\citenamefont {Gietka}\ and\ \citenamefont {Ritsch}(2023)}]{gietka2023squezeingsocbec}%
  \BibitemOpen
  \bibfield  {author} {\bibinfo {author} {\bibfnamefont {K.}~\bibnamefont {Gietka}}\ and\ \bibinfo {author} {\bibfnamefont {H.}~\bibnamefont {Ritsch}},\ }\bibfield  {title} {\bibinfo {title} {Squeezing and {O}vercoming the {H}eisenberg {S}caling with {S}pin-{O}rbit {C}oupled {Q}uantum {G}ases},\ }\href {https://doi.org/10.1103/PhysRevLett.130.090802} {\bibfield  {journal} {\bibinfo  {journal} {Phys. Rev. Lett.}\ }\textbf {\bibinfo {volume} {130}},\ \bibinfo {pages} {090802} (\bibinfo {year} {2023})}\BibitemShut {NoStop}%
\bibitem [{\citenamefont {Salvia}\ \emph {et~al.}(2023)\citenamefont {Salvia}, \citenamefont {Mehboudi},\ and\ \citenamefont {Perarnau-Llobet}}]{marti2023CQMfeedback}%
  \BibitemOpen
  \bibfield  {author} {\bibinfo {author} {\bibfnamefont {R.}~\bibnamefont {Salvia}}, \bibinfo {author} {\bibfnamefont {M.}~\bibnamefont {Mehboudi}},\ and\ \bibinfo {author} {\bibfnamefont {M.}~\bibnamefont {Perarnau-Llobet}},\ }\bibfield  {title} {\bibinfo {title} {Critical {Q}uantum {M}etrology {A}ssisted by {R}eal-{T}ime {F}eedback {C}ontrol},\ }\href {https://doi.org/10.1103/PhysRevLett.130.240803} {\bibfield  {journal} {\bibinfo  {journal} {Phys. Rev. Lett.}\ }\textbf {\bibinfo {volume} {130}},\ \bibinfo {pages} {240803} (\bibinfo {year} {2023})}\BibitemShut {NoStop}%
\bibitem [{\citenamefont {Giannelli}\ \emph {et~al.}(2024)\citenamefont {Giannelli}, \citenamefont {Paladino}, \citenamefont {Grajcar}, \citenamefont {Paraoanu},\ and\ \citenamefont {Falci}}]{Falci2024virtualPhot}%
  \BibitemOpen
  \bibfield  {author} {\bibinfo {author} {\bibfnamefont {L.}~\bibnamefont {Giannelli}}, \bibinfo {author} {\bibfnamefont {E.}~\bibnamefont {Paladino}}, \bibinfo {author} {\bibfnamefont {M.}~\bibnamefont {Grajcar}}, \bibinfo {author} {\bibfnamefont {G.~S.}\ \bibnamefont {Paraoanu}},\ and\ \bibinfo {author} {\bibfnamefont {G.}~\bibnamefont {Falci}},\ }\bibfield  {title} {\bibinfo {title} {Detecting virtual photons in ultrastrongly coupled superconducting quantum circuits},\ }\href {https://doi.org/10.1103/PhysRevResearch.6.013008} {\bibfield  {journal} {\bibinfo  {journal} {Phys. Rev. Res.}\ }\textbf {\bibinfo {volume} {6}},\ \bibinfo {pages} {013008} (\bibinfo {year} {2024})}\BibitemShut {NoStop}%
\bibitem [{\citenamefont {Giovannetti}\ \emph {et~al.}(2006)\citenamefont {Giovannetti}, \citenamefont {Lloyd},\ and\ \citenamefont {Maccone}}]{LLOYD2006QM}%
  \BibitemOpen
  \bibfield  {author} {\bibinfo {author} {\bibfnamefont {V.}~\bibnamefont {Giovannetti}}, \bibinfo {author} {\bibfnamefont {S.}~\bibnamefont {Lloyd}},\ and\ \bibinfo {author} {\bibfnamefont {L.}~\bibnamefont {Maccone}},\ }\bibfield  {title} {\bibinfo {title} {Quantum {M}etrology},\ }\href {https://doi.org/10.1103/PhysRevLett.96.010401} {\bibfield  {journal} {\bibinfo  {journal} {Phys. Rev. Lett.}\ }\textbf {\bibinfo {volume} {96}},\ \bibinfo {pages} {010401} (\bibinfo {year} {2006})}\BibitemShut {NoStop}%
\bibitem [{\citenamefont {Giovannetti}\ \emph {et~al.}(2011)\citenamefont {Giovannetti}, \citenamefont {Lloyd},\ and\ \citenamefont {Maccone}}]{lloyd2011advancesQM}%
  \BibitemOpen
  \bibfield  {author} {\bibinfo {author} {\bibfnamefont {V.}~\bibnamefont {Giovannetti}}, \bibinfo {author} {\bibfnamefont {S.}~\bibnamefont {Lloyd}},\ and\ \bibinfo {author} {\bibfnamefont {L.}~\bibnamefont {Maccone}},\ }\bibfield  {title} {\bibinfo {title} {Advances in quantum metrology},\ }\href {https://doi.org/10.1038/nphoton.2011.35} {\bibfield  {journal} {\bibinfo  {journal} {Nat. Photon.}\ }\textbf {\bibinfo {volume} {5}},\ \bibinfo {pages} {222} (\bibinfo {year} {2011})}\BibitemShut {NoStop}%
\bibitem [{\citenamefont {Pezz\`e}\ \emph {et~al.}(2018)\citenamefont {Pezz\`e}, \citenamefont {Smerzi}, \citenamefont {Oberthaler}, \citenamefont {Schmied},\ and\ \citenamefont {Treutlein}}]{smerzi2018rmp}%
  \BibitemOpen
  \bibfield  {author} {\bibinfo {author} {\bibfnamefont {L.}~\bibnamefont {Pezz\`e}}, \bibinfo {author} {\bibfnamefont {A.}~\bibnamefont {Smerzi}}, \bibinfo {author} {\bibfnamefont {M.~K.}\ \bibnamefont {Oberthaler}}, \bibinfo {author} {\bibfnamefont {R.}~\bibnamefont {Schmied}},\ and\ \bibinfo {author} {\bibfnamefont {P.}~\bibnamefont {Treutlein}},\ }\bibfield  {title} {\bibinfo {title} {Quantum metrology with nonclassical states of atomic ensembles},\ }\href {https://doi.org/10.1103/RevModPhys.90.035005} {\bibfield  {journal} {\bibinfo  {journal} {Rev. Mod. Phys.}\ }\textbf {\bibinfo {volume} {90}},\ \bibinfo {pages} {035005} (\bibinfo {year} {2018})}\BibitemShut {NoStop}%
\bibitem [{\citenamefont {Polino}\ \emph {et~al.}(2020)\citenamefont {Polino}, \citenamefont {Valeri}, \citenamefont {Spagnolo},\ and\ \citenamefont {Sciarrino}}]{photonicQM2020sciarrino}%
  \BibitemOpen
  \bibfield  {author} {\bibinfo {author} {\bibfnamefont {E.}~\bibnamefont {Polino}}, \bibinfo {author} {\bibfnamefont {M.}~\bibnamefont {Valeri}}, \bibinfo {author} {\bibfnamefont {N.}~\bibnamefont {Spagnolo}},\ and\ \bibinfo {author} {\bibfnamefont {F.}~\bibnamefont {Sciarrino}},\ }\bibfield  {title} {\bibinfo {title} {Photonic quantum metrology},\ }\href {https://doi.org/10.1116/5.0007577} {\bibfield  {journal} {\bibinfo  {journal} {AVS Quantum Science}\ }\textbf {\bibinfo {volume} {2}},\ \bibinfo {pages} {024703} (\bibinfo {year} {2020})}\BibitemShut {NoStop}%
\bibitem [{\citenamefont {Pezz\'e}\ and\ \citenamefont {Smerzi}(2009)}]{smerzi2009entangHL}%
  \BibitemOpen
  \bibfield  {author} {\bibinfo {author} {\bibfnamefont {L.}~\bibnamefont {Pezz\'e}}\ and\ \bibinfo {author} {\bibfnamefont {A.}~\bibnamefont {Smerzi}},\ }\bibfield  {title} {\bibinfo {title} {Entanglement, {N}onlinear {D}ynamics, and the {H}eisenberg {L}imit},\ }\href {https://doi.org/10.1103/PhysRevLett.102.100401} {\bibfield  {journal} {\bibinfo  {journal} {Phys. Rev. Lett.}\ }\textbf {\bibinfo {volume} {102}},\ \bibinfo {pages} {100401} (\bibinfo {year} {2009})}\BibitemShut {NoStop}%
\bibitem [{\citenamefont {Demkowicz-Dobrza{\'n}ski}\ \emph {et~al.}(2012)\citenamefont {Demkowicz-Dobrza{\'n}ski}, \citenamefont {Ko{\l}ody{\'n}ski},\ and\ \citenamefont {Gu{\c t}{\u a}}}]{demko2012elusiveHL}%
  \BibitemOpen
  \bibfield  {author} {\bibinfo {author} {\bibfnamefont {R.}~\bibnamefont {Demkowicz-Dobrza{\'n}ski}}, \bibinfo {author} {\bibfnamefont {J.}~\bibnamefont {Ko{\l}ody{\'n}ski}},\ and\ \bibinfo {author} {\bibfnamefont {M.}~\bibnamefont {Gu{\c t}{\u a}}},\ }\bibfield  {title} {\bibinfo {title} {The elusive {H}eisenberg limit in quantum-enhanced metrology},\ }\href {https://doi.org/10.1038/ncomms2067} {\bibfield  {journal} {\bibinfo  {journal} {Nat. Commun.}\ }\textbf {\bibinfo {volume} {3}},\ \bibinfo {pages} {1063} (\bibinfo {year} {2012})}\BibitemShut {NoStop}%
\bibitem [{\citenamefont {Braunstein}\ and\ \citenamefont {Caves}(1994)}]{geometry1994caves}%
  \BibitemOpen
  \bibfield  {author} {\bibinfo {author} {\bibfnamefont {S.~L.}\ \bibnamefont {Braunstein}}\ and\ \bibinfo {author} {\bibfnamefont {C.~M.}\ \bibnamefont {Caves}},\ }\bibfield  {title} {\bibinfo {title} {Statistical distance and the geometry of quantum states},\ }\href {https://doi.org/10.1103/PhysRevLett.72.3439} {\bibfield  {journal} {\bibinfo  {journal} {Phys. Rev. Lett.}\ }\textbf {\bibinfo {volume} {72}},\ \bibinfo {pages} {3439} (\bibinfo {year} {1994})}\BibitemShut {NoStop}%
\bibitem [{\citenamefont {Górecki}\ \emph {et~al.}(2024)\citenamefont {Górecki}, \citenamefont {Albarelli}, \citenamefont {Felicetti}, \citenamefont {Candia},\ and\ \citenamefont {Maccone}}]{górecki2024interplaytimeenergybosonic}%
  \BibitemOpen
  \bibfield  {author} {\bibinfo {author} {\bibfnamefont {W.}~\bibnamefont {Górecki}}, \bibinfo {author} {\bibfnamefont {F.}~\bibnamefont {Albarelli}}, \bibinfo {author} {\bibfnamefont {S.}~\bibnamefont {Felicetti}}, \bibinfo {author} {\bibfnamefont {R.~D.}\ \bibnamefont {Candia}},\ and\ \bibinfo {author} {\bibfnamefont {L.}~\bibnamefont {Maccone}},\ }\href {https://arxiv.org/abs/2409.18791} {\bibinfo {title} {Interplay between time and energy in bosonic noisy quantum metrology}} (\bibinfo {year} {2024}),\ \Eprint {https://arxiv.org/abs/2409.18791} {arXiv:2409.18791 [quant-ph]} \BibitemShut {NoStop}%
\bibitem [{\citenamefont {Oh}\ \emph {et~al.}(2019)\citenamefont {Oh}, \citenamefont {Lee}, \citenamefont {Rockstuhl}, \citenamefont {Jeong}, \citenamefont {Kim}, \citenamefont {Nha},\ and\ \citenamefont {Lee}}]{gaussian2019}%
  \BibitemOpen
  \bibfield  {author} {\bibinfo {author} {\bibfnamefont {C.}~\bibnamefont {Oh}}, \bibinfo {author} {\bibfnamefont {C.}~\bibnamefont {Lee}}, \bibinfo {author} {\bibfnamefont {C.}~\bibnamefont {Rockstuhl}}, \bibinfo {author} {\bibfnamefont {H.}~\bibnamefont {Jeong}}, \bibinfo {author} {\bibfnamefont {J.}~\bibnamefont {Kim}}, \bibinfo {author} {\bibfnamefont {H.}~\bibnamefont {Nha}},\ and\ \bibinfo {author} {\bibfnamefont {S.-Y.}\ \bibnamefont {Lee}},\ }\bibfield  {title} {\bibinfo {title} {Optimal gaussian measurements for phase estimation in single-mode gaussian metrology},\ }\href {https://doi.org/10.1038/s41534-019-0124-4} {\bibfield  {journal} {\bibinfo  {journal} {npj Quantum Inf}\ }\textbf {\bibinfo {volume} {5}},\ \bibinfo {pages} {10} (\bibinfo {year} {2019})}\BibitemShut {NoStop}%
\bibitem [{\citenamefont {Baumann}\ \emph {et~al.}(2010)\citenamefont {Baumann}, \citenamefont {Guerlin}, \citenamefont {Brennecke},\ and\ \citenamefont {Esslinger}}]{esslinger2010dickemodel}%
  \BibitemOpen
  \bibfield  {author} {\bibinfo {author} {\bibfnamefont {K.}~\bibnamefont {Baumann}}, \bibinfo {author} {\bibfnamefont {C.}~\bibnamefont {Guerlin}}, \bibinfo {author} {\bibfnamefont {F.}~\bibnamefont {Brennecke}},\ and\ \bibinfo {author} {\bibfnamefont {T.}~\bibnamefont {Esslinger}},\ }\bibfield  {title} {\bibinfo {title} {Dicke quantum phase transition with a superfluid gas in an optical cavity},\ }\href {https://doi.org/10.1038/nature09009} {\bibfield  {journal} {\bibinfo  {journal} {Nature}\ }\textbf {\bibinfo {volume} {464}},\ \bibinfo {pages} {1301} (\bibinfo {year} {2010})}\BibitemShut {NoStop}%
\bibitem [{\citenamefont {Mottl}\ \emph {et~al.}(2012)\citenamefont {Mottl}, \citenamefont {Brennecke}, \citenamefont {Baumann}, \citenamefont {Landig}, \citenamefont {Donner},\ and\ \citenamefont {Esslinger}}]{esslinger2012rotontype}%
  \BibitemOpen
  \bibfield  {author} {\bibinfo {author} {\bibfnamefont {R.}~\bibnamefont {Mottl}}, \bibinfo {author} {\bibfnamefont {F.}~\bibnamefont {Brennecke}}, \bibinfo {author} {\bibfnamefont {K.}~\bibnamefont {Baumann}}, \bibinfo {author} {\bibfnamefont {R.}~\bibnamefont {Landig}}, \bibinfo {author} {\bibfnamefont {T.}~\bibnamefont {Donner}},\ and\ \bibinfo {author} {\bibfnamefont {T.}~\bibnamefont {Esslinger}},\ }\bibfield  {title} {\bibinfo {title} {Roton-{T}ype {M}ode {S}oftening in a {Q}uantum {G}as with {C}avity-{M}ediated {L}ong-{R}ange {I}nteractions},\ }\href {https://doi.org/10.1126/science.1220314} {\bibfield  {journal} {\bibinfo  {journal} {Science}\ }\textbf {\bibinfo {volume} {336}},\ \bibinfo {pages} {1570} (\bibinfo {year} {2012})}\BibitemShut {NoStop}%
\bibitem [{\citenamefont {Bamba}\ and\ \citenamefont {Ogawa}(2012)}]{Bamba2012}%
  \BibitemOpen
  \bibfield  {author} {\bibinfo {author} {\bibfnamefont {M.}~\bibnamefont {Bamba}}\ and\ \bibinfo {author} {\bibfnamefont {T.}~\bibnamefont {Ogawa}},\ }\bibfield  {title} {\bibinfo {title} {Dissipation and detection of polaritons in the ultrastrong-coupling regime},\ }\href {https://doi.org/10.1103/PhysRevA.86.063831} {\bibfield  {journal} {\bibinfo  {journal} {Phys. Rev. A}\ }\textbf {\bibinfo {volume} {86}},\ \bibinfo {pages} {063831} (\bibinfo {year} {2012})}\BibitemShut {NoStop}%
\bibitem [{\citenamefont {Bamba}\ and\ \citenamefont {Ogawa}(2013)}]{Bamba2013}%
  \BibitemOpen
  \bibfield  {author} {\bibinfo {author} {\bibfnamefont {M.}~\bibnamefont {Bamba}}\ and\ \bibinfo {author} {\bibfnamefont {T.}~\bibnamefont {Ogawa}},\ }\bibfield  {title} {\bibinfo {title} {System-environment coupling derived by {M}axwell's boundary conditions from the weak to the ultrastrong light-matter-coupling regime},\ }\href {https://doi.org/10.1103/PhysRevA.88.013814} {\bibfield  {journal} {\bibinfo  {journal} {Phys. Rev. A}\ }\textbf {\bibinfo {volume} {88}},\ \bibinfo {pages} {013814} (\bibinfo {year} {2013})}\BibitemShut {NoStop}%
\bibitem [{\citenamefont {De~Liberato}(2014)}]{DeLiberato2014}%
  \BibitemOpen
  \bibfield  {author} {\bibinfo {author} {\bibfnamefont {S.}~\bibnamefont {De~Liberato}},\ }\bibfield  {title} {\bibinfo {title} {Comment on ``{S}ystem-environment coupling derived by {M}axwell's boundary conditions from the weak to the ultrastrong light-matter-coupling regime''},\ }\href {https://doi.org/10.1103/PhysRevA.89.017801} {\bibfield  {journal} {\bibinfo  {journal} {Phys. Rev. A}\ }\textbf {\bibinfo {volume} {89}},\ \bibinfo {pages} {017801} (\bibinfo {year} {2014})}\BibitemShut {NoStop}%
\bibitem [{\citenamefont {Barberena}\ \emph {et~al.}(2023)\citenamefont {Barberena}, \citenamefont {Lewis-Swan}, \citenamefont {Rey},\ and\ \citenamefont {Thompson}}]{thompson2023ultranarrow}%
  \BibitemOpen
  \bibfield  {author} {\bibinfo {author} {\bibfnamefont {D.}~\bibnamefont {Barberena}}, \bibinfo {author} {\bibfnamefont {R.~J.}\ \bibnamefont {Lewis-Swan}}, \bibinfo {author} {\bibfnamefont {A.~M.}\ \bibnamefont {Rey}},\ and\ \bibinfo {author} {\bibfnamefont {J.~K.}\ \bibnamefont {Thompson}},\ }\bibfield  {title} {\bibinfo {title} {Ultra narrow linewidth frequency reference via measurement and feedback},\ }\href {https://doi.org/10.5802/crphys.146} {\bibfield  {journal} {\bibinfo  {journal} {Comptes Rendus. Physique}\ }\textbf {\bibinfo {volume} {24}},\ \bibinfo {pages} {55} (\bibinfo {year} {2023})}\BibitemShut {NoStop}%
\bibitem [{\citenamefont {Mivehvar}\ \emph {et~al.}(2021)\citenamefont {Mivehvar}, \citenamefont {Piazza}, \citenamefont {Donner},\ and\ \citenamefont {Ritsch}}]{Mivehvar02012021}%
  \BibitemOpen
  \bibfield  {author} {\bibinfo {author} {\bibfnamefont {F.}~\bibnamefont {Mivehvar}}, \bibinfo {author} {\bibfnamefont {F.}~\bibnamefont {Piazza}}, \bibinfo {author} {\bibfnamefont {T.}~\bibnamefont {Donner}},\ and\ \bibinfo {author} {\bibfnamefont {H.}~\bibnamefont {Ritsch}},\ }\bibfield  {title} {\bibinfo {title} {Cavity {Q}{E}{D} with quantum gases: new paradigms in many-body physics},\ }\href {https://doi.org/10.1080/00018732.2021.1969727} {\bibfield  {journal} {\bibinfo  {journal} {Adv Phys.}\ }\textbf {\bibinfo {volume} {70}},\ \bibinfo {pages} {1} (\bibinfo {year} {2021})}\BibitemShut {NoStop}%
\bibitem [{\citenamefont {Birnbaum}\ \emph {et~al.}(2005)\citenamefont {Birnbaum}, \citenamefont {Boca}, \citenamefont {Miller}, \citenamefont {Boozer}, \citenamefont {Northup},\ and\ \citenamefont {Kimble}}]{kimble2005photonblockade}%
  \BibitemOpen
  \bibfield  {author} {\bibinfo {author} {\bibfnamefont {K.~M.}\ \bibnamefont {Birnbaum}}, \bibinfo {author} {\bibfnamefont {A.}~\bibnamefont {Boca}}, \bibinfo {author} {\bibfnamefont {R.}~\bibnamefont {Miller}}, \bibinfo {author} {\bibfnamefont {A.~D.}\ \bibnamefont {Boozer}}, \bibinfo {author} {\bibfnamefont {T.~E.}\ \bibnamefont {Northup}},\ and\ \bibinfo {author} {\bibfnamefont {H.~J.}\ \bibnamefont {Kimble}},\ }\bibfield  {title} {\bibinfo {title} {Photon blockade in an optical cavity with one trapped atom},\ }\href {https://doi.org/10.1038/nature03804} {\bibfield  {journal} {\bibinfo  {journal} {Nature}\ }\textbf {\bibinfo {volume} {436}},\ \bibinfo {pages} {87} (\bibinfo {year} {2005})}\BibitemShut {NoStop}%
\bibitem [{\citenamefont {Hopfield}(1958)}]{hopfield1958hopfieldmodel}%
  \BibitemOpen
  \bibfield  {author} {\bibinfo {author} {\bibfnamefont {J.~J.}\ \bibnamefont {Hopfield}},\ }\bibfield  {title} {\bibinfo {title} {Theory of the {C}ontribution of {E}xcitons to the {C}omplex {D}ielectric {C}onstant of {C}rystals},\ }\href {https://doi.org/10.1103/PhysRev.112.1555} {\bibfield  {journal} {\bibinfo  {journal} {Phys. Rev.}\ }\textbf {\bibinfo {volume} {112}},\ \bibinfo {pages} {1555} (\bibinfo {year} {1958})}\BibitemShut {NoStop}%
\bibitem [{\citenamefont {Markovi\ifmmode~\acute{c}\else \'{c}\fi{}}\ \emph {et~al.}(2018)\citenamefont {Markovi\ifmmode~\acute{c}\else \'{c}\fi{}}, \citenamefont {Jezouin}, \citenamefont {Ficheux}, \citenamefont {Fedortchenko}, \citenamefont {Felicetti}, \citenamefont {Coudreau}, \citenamefont {Milman}, \citenamefont {Leghtas},\ and\ \citenamefont {Huard}}]{huard2018effectiveUSC}%
  \BibitemOpen
  \bibfield  {author} {\bibinfo {author} {\bibfnamefont {D.}~\bibnamefont {Markovi\ifmmode~\acute{c}\else \'{c}\fi{}}}, \bibinfo {author} {\bibfnamefont {S.}~\bibnamefont {Jezouin}}, \bibinfo {author} {\bibfnamefont {Q.}~\bibnamefont {Ficheux}}, \bibinfo {author} {\bibfnamefont {S.}~\bibnamefont {Fedortchenko}}, \bibinfo {author} {\bibfnamefont {S.}~\bibnamefont {Felicetti}}, \bibinfo {author} {\bibfnamefont {T.}~\bibnamefont {Coudreau}}, \bibinfo {author} {\bibfnamefont {P.}~\bibnamefont {Milman}}, \bibinfo {author} {\bibfnamefont {Z.}~\bibnamefont {Leghtas}},\ and\ \bibinfo {author} {\bibfnamefont {B.}~\bibnamefont {Huard}},\ }\bibfield  {title} {\bibinfo {title} {Demonstration of an {E}ffective {U}ltrastrong {C}oupling {b}etween {T}wo {O}scillators},\ }\href {https://doi.org/10.1103/PhysRevLett.121.040505} {\bibfield  {journal} {\bibinfo  {journal} {Phys. Rev. Lett.}\ }\textbf {\bibinfo {volume} {121}},\ \bibinfo {pages} {040505} (\bibinfo {year} {2018})}\BibitemShut {NoStop}%
\bibitem [{\citenamefont {Kustura}\ \emph {et~al.}(2022)\citenamefont {Kustura}, \citenamefont {Gonzalez-Ballestero}, \citenamefont {Sommer}, \citenamefont {Meyer}, \citenamefont {Quidant},\ and\ \citenamefont {Romero-Isart}}]{oriol2022mechanicalsqueezincUSC}%
  \BibitemOpen
  \bibfield  {author} {\bibinfo {author} {\bibfnamefont {K.}~\bibnamefont {Kustura}}, \bibinfo {author} {\bibfnamefont {C.}~\bibnamefont {Gonzalez-Ballestero}}, \bibinfo {author} {\bibfnamefont {A.~d. l.~R.}\ \bibnamefont {Sommer}}, \bibinfo {author} {\bibfnamefont {N.}~\bibnamefont {Meyer}}, \bibinfo {author} {\bibfnamefont {R.}~\bibnamefont {Quidant}},\ and\ \bibinfo {author} {\bibfnamefont {O.}~\bibnamefont {Romero-Isart}},\ }\bibfield  {title} {\bibinfo {title} {Mechanical {S}queezing via {U}nstable {D}ynamics in a {M}icrocavity},\ }\href {https://doi.org/10.1103/PhysRevLett.128.143601} {\bibfield  {journal} {\bibinfo  {journal} {Phys. Rev. Lett.}\ }\textbf {\bibinfo {volume} {128}},\ \bibinfo {pages} {143601} (\bibinfo {year} {2022})}\BibitemShut {NoStop}%
\bibitem [{\citenamefont {Mihailescu}\ \emph {et~al.}(2024)\citenamefont {Mihailescu}, \citenamefont {Campbell},\ and\ \citenamefont {Gietka}}]{mihailescu2024uncertainquantumcriticalmetrology}%
  \BibitemOpen
\bibfield  {author} {\bibinfo {author} {\bibfnamefont {G.}~\bibnamefont {Mihailescu}},
  \bibinfo {author} {\bibfnamefont {S.}~\bibnamefont {Campbell}},\ and\
  \bibinfo {author} {\bibfnamefont {K.}~\bibnamefont {Gietka}},\ }
  \bibfield  {title} {\bibinfo {title} {Uncertain {Q}uantum {C}ritical {M}etrology:
  {F}rom {S}ingle to {M}ulti {P}arameter {S}ensing},\ } \href {https://doi.org/10.1103/PhysRevA.111.052621}
  {\bibfield  {journal} {\bibinfo  {journal} {Phys. Rev. A}\ }\textbf {\bibinfo {volume} {111}},
  \bibinfo {pages} {052621} (\bibinfo {year} {2025})}\BibitemShut {NoStop}%
\bibitem [{\citenamefont {Hwang}\ \emph {et~al.}(2015)\citenamefont {Hwang}, \citenamefont {Puebla},\ and\ \citenamefont {Plenio}}]{plenio2016qptRM}%
  \BibitemOpen
  \bibfield  {author} {\bibinfo {author} {\bibfnamefont {M.-J.}\ \bibnamefont {Hwang}}, \bibinfo {author} {\bibfnamefont {R.}~\bibnamefont {Puebla}},\ and\ \bibinfo {author} {\bibfnamefont {M.~B.}\ \bibnamefont {Plenio}},\ }\bibfield  {title} {\bibinfo {title} {Quantum {P}hase {T}ransition and {U}niversal {D}ynamics in the {R}abi {M}odel},\ }\href {https://doi.org/10.1103/PhysRevLett.115.180404} {\bibfield  {journal} {\bibinfo  {journal} {Phys. Rev. Lett.}\ }\textbf {\bibinfo {volume} {115}},\ \bibinfo {pages} {180404} (\bibinfo {year} {2015})}\BibitemShut {NoStop}%
\bibitem [{\citenamefont {Gietka}\ and\ \citenamefont {Busch}(2021)}]{gietka2021invertedoscillator}%
  \BibitemOpen
  \bibfield  {author} {\bibinfo {author} {\bibfnamefont {K.}~\bibnamefont {Gietka}}\ and\ \bibinfo {author} {\bibfnamefont {T.}~\bibnamefont {Busch}},\ }\bibfield  {title} {\bibinfo {title} {Inverted harmonic oscillator dynamics of the nonequilibrium phase transition in the {D}icke model},\ }\href {https://doi.org/10.1103/PhysRevE.104.034132} {\bibfield  {journal} {\bibinfo  {journal} {Phys. Rev. E}\ }\textbf {\bibinfo {volume} {104}},\ \bibinfo {pages} {034132} (\bibinfo {year} {2021})}\BibitemShut {NoStop}%
\bibitem [{\citenamefont {Zanardi}\ \emph {et~al.}(2007)\citenamefont {Zanardi}, \citenamefont {Giorda},\ and\ \citenamefont {Cozzini}}]{Zanardi2007Informationgeometry}%
  \BibitemOpen
  \bibfield  {author} {\bibinfo {author} {\bibfnamefont {P.}~\bibnamefont {Zanardi}}, \bibinfo {author} {\bibfnamefont {P.}~\bibnamefont {Giorda}},\ and\ \bibinfo {author} {\bibfnamefont {M.}~\bibnamefont {Cozzini}},\ }\bibfield  {title} {\bibinfo {title} {Information-theoretic differential geometry of quantum phase transitions},\ }\href {https://doi.org/10.1103/PhysRevLett.99.100603} {\bibfield  {journal} {\bibinfo  {journal} {Phys. Rev. Lett.}\ }\textbf {\bibinfo {volume} {99}},\ \bibinfo {pages} {100603} (\bibinfo {year} {2007})}\BibitemShut {NoStop}%
\bibitem [{\citenamefont {H\"ansch}(2006)}]{RevModPhys.78.1297}%
  \BibitemOpen
  \bibfield  {author} {\bibinfo {author} {\bibfnamefont {T.~W.}\ \bibnamefont {H\"ansch}},\ }\bibfield  {title} {\bibinfo {title} {Nobel {L}ecture: {P}assion for precision},\ }\href {https://doi.org/10.1103/RevModPhys.78.1297} {\bibfield  {journal} {\bibinfo  {journal} {Rev. Mod. Phys.}\ }\textbf {\bibinfo {volume} {78}},\ \bibinfo {pages} {1297} (\bibinfo {year} {2006})}\BibitemShut {NoStop}%
\bibitem [{\citenamefont {Hotter}\ \emph {et~al.}(2024)\citenamefont {Hotter}, \citenamefont {Ritsch},\ and\ \citenamefont {Gietka}}]{hotter2024combining}%
  \BibitemOpen
  \bibfield  {author} {\bibinfo {author} {\bibfnamefont {C.}~\bibnamefont {Hotter}}, \bibinfo {author} {\bibfnamefont {H.}~\bibnamefont {Ritsch}},\ and\ \bibinfo {author} {\bibfnamefont {K.}~\bibnamefont {Gietka}},\ }\bibfield  {title} {\bibinfo {title} {Combining {C}ritical and {Q}uantum {M}etrology},\ }\href {https://doi.org/10.1103/PhysRevLett.132.060801} {\bibfield  {journal} {\bibinfo  {journal} {Phys. Rev. Lett.}\ }\textbf {\bibinfo {volume} {132}},\ \bibinfo {pages} {060801} (\bibinfo {year} {2024})}\BibitemShut {NoStop}%
\bibitem [{\citenamefont {Klinder}\ \emph {et~al.}(2015)\citenamefont {Klinder}, \citenamefont {Keßler}, \citenamefont {Wolke}, \citenamefont {Mathey},\ and\ \citenamefont {Hemmerich}}]{hemmerich2015dickedynamic}%
  \BibitemOpen
  \bibfield  {author} {\bibinfo {author} {\bibfnamefont {J.}~\bibnamefont {Klinder}}, \bibinfo {author} {\bibfnamefont {H.}~\bibnamefont {Keßler}}, \bibinfo {author} {\bibfnamefont {M.}~\bibnamefont {Wolke}}, \bibinfo {author} {\bibfnamefont {L.}~\bibnamefont {Mathey}},\ and\ \bibinfo {author} {\bibfnamefont {A.}~\bibnamefont {Hemmerich}},\ }\bibfield  {title} {\bibinfo {title} {Dynamical phase transition in the open {D}icke model},\ }\href {https://doi.org/10.1073/pnas.1417132112} {\bibfield  {journal} {\bibinfo  {journal} {Proc. Natl. Acad. Sci. U.S.A.}\ }\textbf {\bibinfo {volume} {112}},\ \bibinfo {pages} {3290} (\bibinfo {year} {2015})}\BibitemShut {NoStop}%
\bibitem [{\citenamefont {Koll{\'a}r}\ \emph {et~al.}(2017)\citenamefont {Koll{\'a}r}, \citenamefont {Papageorge}, \citenamefont {Vaidya}, \citenamefont {Guo}, \citenamefont {Keeling},\ and\ \citenamefont {Lev}}]{lev2017phasetransitionpolariton}%
  \BibitemOpen
  \bibfield  {author} {\bibinfo {author} {\bibfnamefont {A.~J.}\ \bibnamefont {Koll{\'a}r}}, \bibinfo {author} {\bibfnamefont {A.~T.}\ \bibnamefont {Papageorge}}, \bibinfo {author} {\bibfnamefont {V.~D.}\ \bibnamefont {Vaidya}}, \bibinfo {author} {\bibfnamefont {Y.}~\bibnamefont {Guo}}, \bibinfo {author} {\bibfnamefont {J.}~\bibnamefont {Keeling}},\ and\ \bibinfo {author} {\bibfnamefont {B.~L.}\ \bibnamefont {Lev}},\ }\bibfield  {title} {\bibinfo {title} {Supermode-density-wave-polariton condensation with a {B}ose--{E}instein condensate in a multimode cavity},\ }\href {https://doi.org/10.1038/ncomms14386} {\bibfield  {journal} {\bibinfo  {journal} {Nat. Commun.}\ }\textbf {\bibinfo {volume} {8}},\ \bibinfo {pages} {14386} (\bibinfo {year} {2017})}\BibitemShut {NoStop}%
\bibitem [{\citenamefont {Yoshihara}\ \emph {et~al.}(2017)\citenamefont {Yoshihara}, \citenamefont {Fuse}, \citenamefont {Ashhab}, \citenamefont {Kakuyanagi}, \citenamefont {Saito},\ and\ \citenamefont {Semba}}]{Yoshihara2016}%
  \BibitemOpen
  \bibfield  {author} {\bibinfo {author} {\bibfnamefont {F.}~\bibnamefont {Yoshihara}}, \bibinfo {author} {\bibfnamefont {T.}~\bibnamefont {Fuse}}, \bibinfo {author} {\bibfnamefont {S.}~\bibnamefont {Ashhab}}, \bibinfo {author} {\bibfnamefont {K.}~\bibnamefont {Kakuyanagi}}, \bibinfo {author} {\bibfnamefont {S.}~\bibnamefont {Saito}},\ and\ \bibinfo {author} {\bibfnamefont {K.}~\bibnamefont {Semba}},\ }\bibfield  {title} {\bibinfo {title} {Superconducting qubit--oscillator circuit beyond the ultrastrong-coupling regime},\ }\href {https://doi.org/10.1038/nphys3906} {\bibfield  {journal} {\bibinfo  {journal} {Nature Physics}\ }\textbf {\bibinfo {volume} {13}},\ \bibinfo {pages} {44} (\bibinfo {year} {2017})}\BibitemShut {NoStop}%
\bibitem [{\citenamefont {Gu}\ \emph {et~al.}(2017)\citenamefont {Gu}, \citenamefont {Kockum}, \citenamefont {Miranowicz}, \citenamefont {xi~Liu},\ and\ \citenamefont {Nori}}]{Gu2017}%
  \BibitemOpen
  \bibfield  {author} {\bibinfo {author} {\bibfnamefont {X.}~\bibnamefont {Gu}}, \bibinfo {author} {\bibfnamefont {A.~F.}\ \bibnamefont {Kockum}}, \bibinfo {author} {\bibfnamefont {A.}~\bibnamefont {Miranowicz}}, \bibinfo {author} {\bibfnamefont {Y.}~\bibnamefont {xi~Liu}},\ and\ \bibinfo {author} {\bibfnamefont {F.}~\bibnamefont {Nori}},\ }\bibfield  {title} {\bibinfo {title} {Microwave photonics with superconducting quantum circuits},\ }\href {https://doi.org/https://doi.org/10.1016/j.physrep.2017.10.002} {\bibfield  {journal} {\bibinfo  {journal} {Physics Reports}\ }\textbf {\bibinfo {volume} {718-719}},\ \bibinfo {pages} {1} (\bibinfo {year} {2017})},\ \bibinfo {note} {microwave photonics with superconducting quantum circuits}\BibitemShut {NoStop}%
\bibitem [{\citenamefont {Dare}\ \emph {et~al.}(2024{\natexlab{b}})\citenamefont {Dare}, \citenamefont {Hansen}, \citenamefont {Coroli}, \citenamefont {Johnson}, \citenamefont {Aspelmeyer},\ and\ \citenamefont {Deli\ifmmode~\acute{c}\else \'{c}\fi{}}}]{delic2024USCoptomechanics}%
  \BibitemOpen
  \bibfield  {author} {\bibinfo {author} {\bibfnamefont {K.}~\bibnamefont {Dare}}, \bibinfo {author} {\bibfnamefont {J.~J.}\ \bibnamefont {Hansen}}, \bibinfo {author} {\bibfnamefont {I.}~\bibnamefont {Coroli}}, \bibinfo {author} {\bibfnamefont {A.}~\bibnamefont {Johnson}}, \bibinfo {author} {\bibfnamefont {M.}~\bibnamefont {Aspelmeyer}},\ and\ \bibinfo {author} {\bibfnamefont {U.}~\bibnamefont {Deli\ifmmode~\acute{c}\else \'{c}\fi{}}},\ }\bibfield  {title} {\bibinfo {title} {Ultrastrong linear optomechanical interaction},\ }\href {https://doi.org/10.1103/PhysRevResearch.6.L042025} {\bibfield  {journal} {\bibinfo  {journal} {Phys. Rev. Res.}\ }\textbf {\bibinfo {volume} {6}},\ \bibinfo {pages} {L042025} (\bibinfo {year} {2024}{\natexlab{b}})}\BibitemShut {NoStop}%
\bibitem [{\citenamefont {Braum{\"u}ller}\ \emph {et~al.}(2017)\citenamefont {Braum{\"u}ller}, \citenamefont {Marthaler}, \citenamefont {Schneider}, \citenamefont {Stehli}, \citenamefont {Rotzinger}, \citenamefont {Weides},\ and\ \citenamefont {Ustinov}}]{Ustinow2017USC_QRM}%
  \BibitemOpen
  \bibfield  {author} {\bibinfo {author} {\bibfnamefont {J.}~\bibnamefont {Braum{\"u}ller}}, \bibinfo {author} {\bibfnamefont {M.}~\bibnamefont {Marthaler}}, \bibinfo {author} {\bibfnamefont {A.}~\bibnamefont {Schneider}}, \bibinfo {author} {\bibfnamefont {A.}~\bibnamefont {Stehli}}, \bibinfo {author} {\bibfnamefont {H.}~\bibnamefont {Rotzinger}}, \bibinfo {author} {\bibfnamefont {M.}~\bibnamefont {Weides}},\ and\ \bibinfo {author} {\bibfnamefont {A.~V.}\ \bibnamefont {Ustinov}},\ }\bibfield  {title} {\bibinfo {title} {Analog quantum simulation of the {R}abi model in the ultra-strong coupling regime},\ }\href {https://doi.org/10.1038/s41467-017-00894-w} {\bibfield  {journal} {\bibinfo  {journal} {Nat. Commun.}\ }\textbf {\bibinfo {volume} {8}},\ \bibinfo {pages} {779} (\bibinfo {year} {2017})}\BibitemShut {NoStop}%
\bibitem [{\citenamefont {Safavi-Naini}\ \emph {et~al.}(2018)\citenamefont {Safavi-Naini}, \citenamefont {Lewis-Swan}, \citenamefont {Bohnet}, \citenamefont {G\"arttner}, \citenamefont {Gilmore}, \citenamefont {Jordan}, \citenamefont {Cohn}, \citenamefont {Freericks}, \citenamefont {Rey},\ and\ \citenamefont {Bollinger}}]{rey2018dickeion}%
  \BibitemOpen
  \bibfield  {author} {\bibinfo {author} {\bibfnamefont {A.}~\bibnamefont {Safavi-Naini}}, \bibinfo {author} {\bibfnamefont {R.~J.}\ \bibnamefont {Lewis-Swan}}, \bibinfo {author} {\bibfnamefont {J.~G.}\ \bibnamefont {Bohnet}}, \bibinfo {author} {\bibfnamefont {M.}~\bibnamefont {G\"arttner}}, \bibinfo {author} {\bibfnamefont {K.~A.}\ \bibnamefont {Gilmore}}, \bibinfo {author} {\bibfnamefont {J.~E.}\ \bibnamefont {Jordan}}, \bibinfo {author} {\bibfnamefont {J.}~\bibnamefont {Cohn}}, \bibinfo {author} {\bibfnamefont {J.~K.}\ \bibnamefont {Freericks}}, \bibinfo {author} {\bibfnamefont {A.~M.}\ \bibnamefont {Rey}},\ and\ \bibinfo {author} {\bibfnamefont {J.~J.}\ \bibnamefont {Bollinger}},\ }\bibfield  {title} {\bibinfo {title} {Verification of a {M}any-{I}on {S}imulator of the {D}icke {M}odel {T}hrough {S}low {Q}uenches across a {P}hase {T}ransition},\ }\href {https://doi.org/10.1103/PhysRevLett.121.040503} {\bibfield  {journal} {\bibinfo  {journal} {Phys. Rev. Lett.}\ }\textbf {\bibinfo {volume} {121}},\ \bibinfo
  {pages} {040503} (\bibinfo {year} {2018})}\BibitemShut {NoStop}%
\bibitem [{\citenamefont {Gilmore}\ \emph {et~al.}(2021)\citenamefont {Gilmore}, \citenamefont {Affolter}, \citenamefont {Lewis-Swan}, \citenamefont {Barberena}, \citenamefont {Jordan}, \citenamefont {Rey},\ and\ \citenamefont {Bollinger}}]{bollinger2021sensingdicke}%
  \BibitemOpen
  \bibfield  {author} {\bibinfo {author} {\bibfnamefont {K.~A.}\ \bibnamefont {Gilmore}}, \bibinfo {author} {\bibfnamefont {M.}~\bibnamefont {Affolter}}, \bibinfo {author} {\bibfnamefont {R.~J.}\ \bibnamefont {Lewis-Swan}}, \bibinfo {author} {\bibfnamefont {D.}~\bibnamefont {Barberena}}, \bibinfo {author} {\bibfnamefont {E.}~\bibnamefont {Jordan}}, \bibinfo {author} {\bibfnamefont {A.~M.}\ \bibnamefont {Rey}},\ and\ \bibinfo {author} {\bibfnamefont {J.~J.}\ \bibnamefont {Bollinger}},\ }\bibfield  {title} {\bibinfo {title} {Quantum-enhanced sensing of displacements and electric fields with two-dimensional trapped-ion crystals},\ }\href {https://doi.org/10.1126/science.abi5226} {\bibfield  {journal} {\bibinfo  {journal} {Science}\ }\textbf {\bibinfo {volume} {373}},\ \bibinfo {pages} {673} (\bibinfo {year} {2021})}\BibitemShut {NoStop}%
\bibitem [{\citenamefont {Cai}\ \emph {et~al.}(2021)\citenamefont {Cai}, \citenamefont {Liu}, \citenamefont {Zhao}, \citenamefont {Wu}, \citenamefont {Mei}, \citenamefont {Jiang}, \citenamefont {He}, \citenamefont {Zhang}, \citenamefont {Zhou},\ and\ \citenamefont {Duan}}]{duan2021qrmsingleion}%
  \BibitemOpen
  \bibfield  {author} {\bibinfo {author} {\bibfnamefont {M.~L.}\ \bibnamefont {Cai}}, \bibinfo {author} {\bibfnamefont {Z.~D.}\ \bibnamefont {Liu}}, \bibinfo {author} {\bibfnamefont {W.~D.}\ \bibnamefont {Zhao}}, \bibinfo {author} {\bibfnamefont {Y.~K.}\ \bibnamefont {Wu}}, \bibinfo {author} {\bibfnamefont {Q.~X.}\ \bibnamefont {Mei}}, \bibinfo {author} {\bibfnamefont {Y.}~\bibnamefont {Jiang}}, \bibinfo {author} {\bibfnamefont {L.}~\bibnamefont {He}}, \bibinfo {author} {\bibfnamefont {X.}~\bibnamefont {Zhang}}, \bibinfo {author} {\bibfnamefont {Z.~C.}\ \bibnamefont {Zhou}},\ and\ \bibinfo {author} {\bibfnamefont {L.~M.}\ \bibnamefont {Duan}},\ }\bibfield  {title} {\bibinfo {title} {Observation of a quantum phase transition in the quantum {R}abi model with a single trapped ion},\ }\href {https://doi.org/10.1038/s41467-021-21425-8} {\bibfield  {journal} {\bibinfo  {journal} {Nat. Commun.}\ }\textbf {\bibinfo {volume} {12}},\ \bibinfo {pages} {1126} (\bibinfo {year} {2021})}\BibitemShut {NoStop}%
\bibitem [{\citenamefont {Krämer}\ \emph {et~al.}(2018)\citenamefont {Krämer}, \citenamefont {Plankensteiner}, \citenamefont {Ostermann},\ and\ \citenamefont {Ritsch}}]{kramer2018quantumoptics}%
  \BibitemOpen
  \bibfield  {author} {\bibinfo {author} {\bibfnamefont {S.}~\bibnamefont {Krämer}}, \bibinfo {author} {\bibfnamefont {D.}~\bibnamefont {Plankensteiner}}, \bibinfo {author} {\bibfnamefont {L.}~\bibnamefont {Ostermann}},\ and\ \bibinfo {author} {\bibfnamefont {H.}~\bibnamefont {Ritsch}},\ }\bibfield  {title} {\bibinfo {title} {Quantumoptics.jl: A {J}ulia framework for simulating open quantum systems},\ }\href {https://doi.org/https://doi.org/10.1016/j.cpc.2018.02.004} {\bibfield  {journal} {\bibinfo  {journal} {Comput. Phys. Commun.}\ }\textbf {\bibinfo {volume} {227}},\ \bibinfo {pages} {109} (\bibinfo {year} {2018})}\BibitemShut {NoStop}%
\end{thebibliography}
\end{document}